\documentclass{aa}
\usepackage{graphicx}
\usepackage{amsmath}	
\usepackage{siunitx}    
\usepackage{array}
\usepackage{multirow}
\usepackage{booktabs}
\usepackage{longtable}
\usepackage{lscape}

\newcommand{\laedd}{$\lambda_{\mathrm{Edd}}$}

\newcommand{\mbh}{$M_\mathrm{BH}$}

\newcommand{\msigma}{$M\text{--}\sigma_\star$}

\newcommand{\msun}{M$_{\sun}$}

\newcommand{\nh}{$N_{\mathrm{H}}$}

\newcommand{\xmm}{{\it XMM-Newton}}
\newcommand{\xspec}{\textsc{xspec}}

\usepackage{txfonts}
\begin{document}

   \title{Measuring monster MBHs: maybe mighty, maybe merely massive}

   \author{M. Gliozzi
          \inst{1}, A. Akylas \inst{2}, J. K. Williams \inst{1},
          \and I. E. Papadakis
          \inst{3}
          }

   \institute{Department of Physics and Astronomy, George Mason University, 4400 University Drive, Fairfax, VA 22030, USA\\
              \email{mgliozzi@gmu.edu}
         \and
             Institute for Astronomy Astrophysics Space Applications and Remote Sensing (IAASARS), National Observatory of Athens,
I. Metaxa \& V. Pavlou, Penteli 15236, Greece
            \and
            Physics Department, University of Crete, 73010, Heraklion, Greece
             }

   \date{Received ; accepted}

 
  \abstract
   {Accurate black hole mass (\mbh) measurements in high-redshift galaxies are difficult yet crucial to constrain the growth of supermassive BHs within their host galaxies, and to discriminate between competing BH seed models. Recent studies claimed the detection of massive BHs in very distant active galactic nuclei (AGN), implying extreme conditions with massive BH seeds accreting at high rates for extensive periods of time. However, these estimates are usually obtained by extrapolating indirect methods that are calibrated for moderately accreting, low-luminosity AGN in the local universe.}
   {We want to assess the reliability of optically based indirect methods, more specifically the single epoch method (SE), in the distant universe.}
   {We compute the \mbh\ values for a sample of hyperluminous distant quasars (the X-WISSH sample) and a sample of highly accreting AGN (X-HESS) using the X-ray scaling method, which provides values fully consistent with those obtained with direct dynamical methods in the local universe.}
   {We first verify that the X-ray scaling method yields reliable \mbh\ values also for distant highly accreting objects. Then, we carry out a systematic comparison with the SE method and find that these two indirect methods, which are based on completely different assumptions, are fully consistent with each other over a broad range of luminosities, intrinsic absorption, and accretion rates (excluding the SE values based on \ion{C}{iv}). The only discrepancies are associated with AGN that are substantially absorbed, whose \mbh\ are underestimated by the SE method, and AGN accreting well above the Eddington limit, whose \mbh\ appear to be overestimated by the SE method. The latter result casts some doubts on the claim of overmassive BHs estimated with the SE method for highly accreting AGN in the early universe. Our study also reveals that one of the frequently used SDSS AGN catalogs consistently underestimates the \mbh\ values by a factor of 2.5. Although this factor is of the order of the uncertainty generally associated with the SE method, we demonstrate that the use of underestimated values may result in potentially misleading conclusions. Specifically, for this AGN sample we confirm strong positive correlations for the photon index $\Gamma$ vs. the Eddington ratio \laedd\ and for the X-ray bolometric correction vs. \laedd, as well as for $\Gamma$ vs. the soft excess strength, at odds with the conclusions inferred using underestimated \mbh\ values.}
   {}
   \keywords{Black hole physics --
                Galaxies: active --
                X-rays: galaxies
               }

   \maketitle
%

\section{Introduction}
It is now widely accepted that every massive galaxy hosts a supermassive black hole (SMBH) at its center, and that there exists a positive correlation between the mass of the BH (\mbh) and that of the galaxy bulge, suggesting a coevolution between these two components \citep{Magorrian1998,Ferrarese2000,Gebhardt2000}.

Recent discoveries of quasars with \mbh\ values of the order of $10^9\textrm{--}10^{10}\,\mathrm{M_\odot}$ at redshifts of $\sim 7$ \citep{Mortlock2011,Venemans2013,Banados2018,Fan2023} and, more recently, the advent of the James Webb Space Telescope (JWST) led to the detections of numerous galaxies at even larger redshifts (the so-called little red dots, LRDs), whose nature is still debated: they may be massive, compact star-forming galaxies \citep{Labbe2023} or low-mass galaxies hosting moderately luminous active galactic nuclei (AGN) with over-massive BHs \citep{Harikane2023,Maiolino2024}. Importantly, very distant LRDs with estimated masses of SMBHs in the  $10^6\textrm{--}10^8\,\mathrm{M_\odot}$ range \citep{Larson2023,Matthee2024,Greene2024,Maiolino2024} impose severe constraints on current models looking for viable pathways of SMBH formation (see \citealt{Regan2024} for a recent review).

In principle, the detection of galaxies and SMBHs over a very large range of redshifts should make it possible to study the evolution of the relationship between these two components over cosmic time. However, accurate measurements of the SMBH mass and galaxy properties become very challenging at high redshifts. Here, we focus on the reliability of \mbh\ measurements in distant AGN. 

In nearby galaxies, where the BH sphere of influence is spatially resolved by current instrumentation, the \mbh\ can be accurately estimated with direct methods by measuring the kinematics of the gas or stellar components in the inner region of weakly active galaxies (e.g., \citealt{ Macchetto1997,Gebhardt2003}) or the kinematics of the broad line region (BLR) in nearby AGN \citep{Gravity2018,Gravity2021}. For more distant AGN, which exhibit correlated variability between the optical-UV continuum emitted from the accretion disk and the line emission from the BLR, the \mbh\ is accurately measured using the reverberation mapping (RM) technique (e.g., \citealt{Blandford1982, Peterson2004}). However, this time- and instrument-intensive method is restricted to objects that vary substantially on a reasonably short time interval, limiting the direct measurements of \mbh\ to moderately luminous nearby AGN. 

For the vast majority of AGN one must rely on indirect methods to estimate the \mbh. For example, a positive correlation between \mbh\ and the velocity dispersion of the galaxy bulge (the so-called \msigma\ correlation), observed in local nearly quiescent galaxies, is often utilized in cases where dynamical methods are not accessible (see \citealt{Kormendy2013} for a comprehensive review). This specific method, however, cannot be used for distant AGN for the following reasons: 1) there is evidence that AGN do not follow the \msigma\ correlation obtained using local inactive galaxies \citep{Caglar2023}; 2) the local \msigma\ correlation has a tendency to systematically overestimate \mbh\ in AGN regardless of the level of obscuration \citep{Gliozzi2024}; 3) since the ultimate goal is to investigate the evolution of the correlation between BH and galaxy properties, one cannot use an \mbh\ estimate inferred from this relationship.

Another indirect method, which is routinely applied to all type-1 AGN (i.e., objects with detectable BLRs), is the single epoch (SE) method, which exploits the tight correlation observed between the size of the BLR $R_\mathrm{BLR}$ and the optical luminosity $L_\mathrm{opt}$ (e.g., \citealt{Kaspi2005, Bentz2013}). The major advantage of the SE method is its ability to estimate \mbh\ based on a single spectrum, which has led to the estimates of several thousands of \mbh\ values at all redshifts utilizing large AGN spectral catalogs (e.g., \citealt{Rakshit2020}). 

However, recent studies have outlined the tendency of this method to overestimate \mbh\ when applied to highly accreting objects (see, e.g., \citealt{Du2015,Du2018,Martinez2019}) and to luminous AGN \citep{Woo2024} since the method was calibrated using nearby, moderately accreting AGN with relatively low luminosities. Additional concerns about the reliability of the SE method applied to high-redshift AGN were cast by very recent studies from \citet{Bertemes2024}, who demonstrated that different optical and UV broad lines may yield substantially different estimates of \mbh\ for the same object, and by \citet{Fries2024}, who performed a velocity-resolved reverberation mapping study of a highly variable quasar over a time interval of 10 years and revealed that the virial product (the basis of the SE method) was inconsistent over time. 

Despite these concerns, the SE method often offers the only option to estimate \mbh\ in distant AGN. Therefore, it is important to assess its reliability and its potential biases when applied to distant quasars and highly accreting AGN. 

In our recent work, using a representative volume-limited sample of hard-X-ray-selected AGN, we carried out a systematic comparison of indirect methods to estimate \mbh\ in all AGN regardless of their level of obscuration. Our analysis demonstrates that X-ray-based methods (specifically, the X-ray scaling method and the variability method based on the excess variance) yield \mbh\ values that are fully consistent with those obtained with dynamical methods, whereas other methods such as the fundamental plane for black hole activity (e.g., \citealt{Merloni2003,Gueltekin2019})  are either unreliable or have the tendency to overestimate \mbh\ (e.g., the \msigma\ correlation for inactive galaxies). Additionally, our study showed that, for broad-lined AGN, the X-ray-based methods were consistent with the SE method based on the H$\alpha$ line \citep{Gliozzi2024}.

In principle, the X-ray-based methods could be extended to distant AGN without any substantial modification, but they may be affected by inherent difficulties. For example, the variability method (e.g., \citealt{Papadakis2004, Ponti2012, Akylas2022}), which is model-independent and hence applicable to all variable AGN, is severely limited by the lack of light curves sufficiently long and with adequate signal-to-noise ratio. On the other hand, it is easier to obtain X-ray spectra of sufficient quality to apply the X-ray scaling method. This method is based on the assumption that the central engine (disk + corona) producing the X-rays works similarly in all black hole systems accreting at moderate or high rate (see \citealt{Shaposhnikov2009,Gliozzi2011} for further details). This is supported by several studies on the similarities between AGN and stellar mass BHs (see, e.g., \citealt{Done2005,Koerding2006,Mchardy2006}), as well as by the tight correlations observed between $\alpha_\mathrm{ox}$ and $L_{2500~\text{\AA}}$ \citep{Steffen2006} and between the X-ray and UV luminosities \citep{Risaliti2015}, which appear to remain unchanged up to z = 6. 

Nevertheless, since the X-ray scaling method,
with the exception of the recent work from \citet{Elias2024}, has only been applied to relatively nearby AGN \citep{Gliozzi2011,Gliozzi2021,Gliozzi2024,Giacche2014,Seifina2018a,Seifina2018b,Shuvo2022,Williams2023}, we must first perform a sanity check and verify that it yields reasonable values also for distant objects. This is accomplished by comparing the \mbh\ values obtained from this method with those of luminous quasars accreting at the Eddington level and with well-defined spectral energy distributions (SEDs). We then carry out a systematic comparison between the SE method and the X-ray scaling one using a large sample of highly accreting X-ray-bright AGN. Finally, we outline the implications of using the X-ray-based \mbh\ as opposed to values obtained with an automated analysis with the SE method by investigating different correlations between various AGN parameters.

The structure of the paper is as follows. In Section 2, we describe the sample selection, as well as the data reduction and analysis of the \xmm\ data. In Section 3, we report the results obtained from the X-ray scaling method and the systematic comparison with the SE method. We  discuss the main findings in Section 4 and draw our conclusions in Section 5. In this paper, we adopt a cosmology with $H_0= 70~\mathrm{km\,s^{-1}\,Mpc^{-1}}$, $\Omega_\mathrm{m}=0.28$, $\Omega_\mathrm{\Lambda}=0.72$, based on the 9-year measurements of the Wilkinson Microwave Anisotropy Probe (WMAP).


\section{Observations}
\label{sec:Observations}
\subsection{Samples Description}
\label{sec:sample} 
In this study we utilized two different samples. First, we focused on the hyperluminous quasars of the WISSH
 sample, which contains objects with $L_\mathrm{bol} > 2\times 10^{47}~\mathrm{erg~s}^{-1}$ obtained from the cross-correlation of the WISE and SDSS catalogs  \citep{Bischetti2017}. The X-ray properties of a subsample of 41 quasars, the X-WISSH sample, were studied by \citet{Martocchia2017}. Of these quasars, 40 sources have \mbh\ values estimated with the SE method based on the \ion{C}{iv} line, and 14 have \mbh\ values based on the H$\beta$ line. Importantly, 35 of these quasars have broadband SEDs, which were used by \citet{Duras2020} to estimate the bolometric luminosity. Here, we restrict our work to 12 objects that have \xmm\ spectral data of sufficient quality to constrain the main parameters of a Comptonization model, which are needed to estimate \mbh\ with the X-ray scaling method. The basic characteristics of this sample of 12 WISSH quasars can be summarized as follows: the redshift $z$ ranges from 2.04 to 4.11 with an average of 2.92, the bolometric luminosity is between $2.8\times10^{47}\,\mathrm{erg \,s^{-1}}$ and $5.0\times10^{48}\,\mathrm{erg \,s^{-1}}$ with an average of $1.0\times10^{48}\,\mathrm{erg \,s^{-1}}$, the 2--10 keV luminosity $L_\mathrm{X}$ ranges from $10^{45}\,\mathrm{erg \,s^{-1}}$ to $1.5\times10^{46}\,\mathrm{erg \,s^{-1}}$ with an average of $5.2\times10^{45}\,\mathrm{erg \,s^{-1}}$,
and the Eddington ratio $\lambda_\mathrm{Edd}=L_\mathrm{bol}/L_\mathrm{Edd}$ has minimum, maximum, and average of 0.08, 3.28, and 0.97, respectively. Throughout the paper, unless otherwise stated, the values of $\lambda_\mathrm{Edd}$ are computed assuming the \mbh\ is obtained with the X-ray scaling method.

Second, we worked on the \xmm\ High-Eddington Serendipitous AGN Sample (X-HESS), recently selected by \citet{Laurenti2024}, which comprises 60 allegedly highly accreting objects with \mbh\ estimated via the SE method with the H$\beta$ line. Of these, we were able to estimate  \mbh\ for 50 objects with the X-ray scaling method. The main characteristics of this second sample are the following: $z$ varies between 0.06 and 3.31 with an average of 1.09, $L_\mathrm{bol}$ ranges from $5\times10^{44}\,\mathrm{erg \,s^{-1}}$ to $2\times10^{48}\,\mathrm{erg \,s^{-1}}$ with an average of $2.9\times10^{46}\,\mathrm{erg \,s^{-1}}$, $L_\mathrm{X}$ is between $4.6\times10^{42}\,\mathrm{erg \,s^{-1}}$ and $1.5\times10^{46}\,\mathrm{erg \,s^{-1}}$ with an average of $1.4\times10^{45}\,\mathrm{erg \,s^{-1}}$, and $\lambda_\mathrm{Edd}$ ranges between 0.01 and 9.7 with an average of 0.7.

The main properties of the objects analyzed in this work are summarized in two tables in the Appendix.

\subsection{Data reduction}
\label{sec:data}
We processed the entire sample using the XMM-Newton pipeline spectral data products available in the XMM-Newton Science Archive (XSA) at the European Space Astronomy Centre (ESAC).
The source spectrum was accumulated by using a spatial filter (a circular aperture whose radius was determined by an S/N optimization algorithm) on valid events in an exposure from CCDs operating in IMAGING mode. For each candidate source, a spectrum was produced for each EPIC camera (two MOS and one pn), where available.
Additionally, for each source spectrum, a background spectrum was produced by accumulating detected events from a source-free region of the field of view (contaminating source regions having been masked out in the process). The corresponding ancillary response function file, which provides the effective area of the instrument as a function of energy, was also produced for each source and exposure for which spectral products have been extracted and was used with those spectral products.

\subsection{X-ray spectral analysis}
\label{sec:X-ray spectral}
The X-ray scaling method is based on the spectral fitting of the primary X-ray emission of the objects in our sample with the bulk Comptonization model \texttt{BMC} \citep{Titarchuk1997}.
To fit the hard X-ray spectrum where the lower energy limit is fixed at 2 keV in the source rest frame, we used a simple baseline model  
\begin{verbatim} 
constant*zphabs*BMC 
\end{verbatim}
where the constant takes into account the difference in calibration among the three \xmm\ EPIC cameras, the absorption model \texttt{zphabs} describes both the Galaxy and the intrinsic contributions, and \texttt{BMC} parameterizes the Comptonization process. 

We performed the X-ray spectral analysis using the \xspec\ \texttt{v.12.9.0} software package \citep{Arnaud1996}, and  the errors quoted on the spectral parameters represent the 1$\sigma$ confidence level.

\begin{figure*}
  \includegraphics[width=0.32\linewidth]{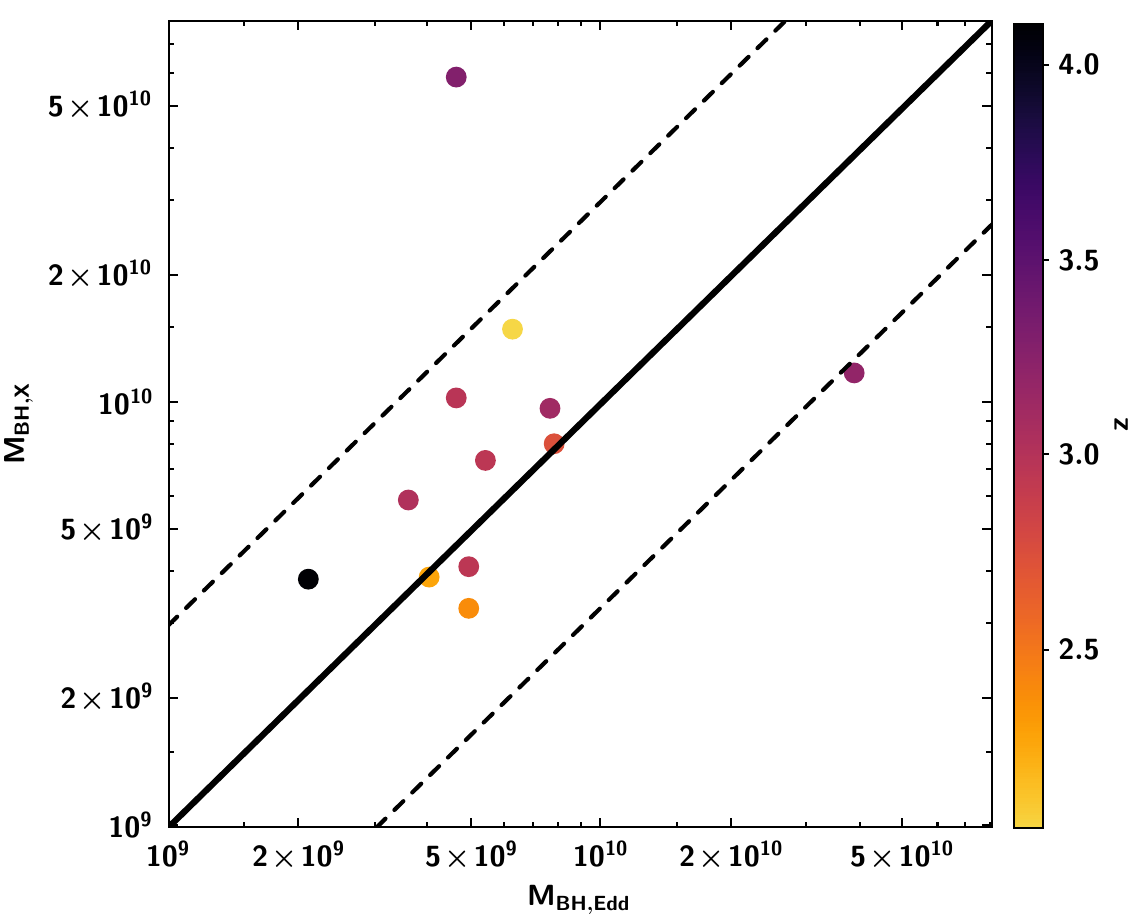}
  \hfill
  \includegraphics[width=0.32\linewidth]{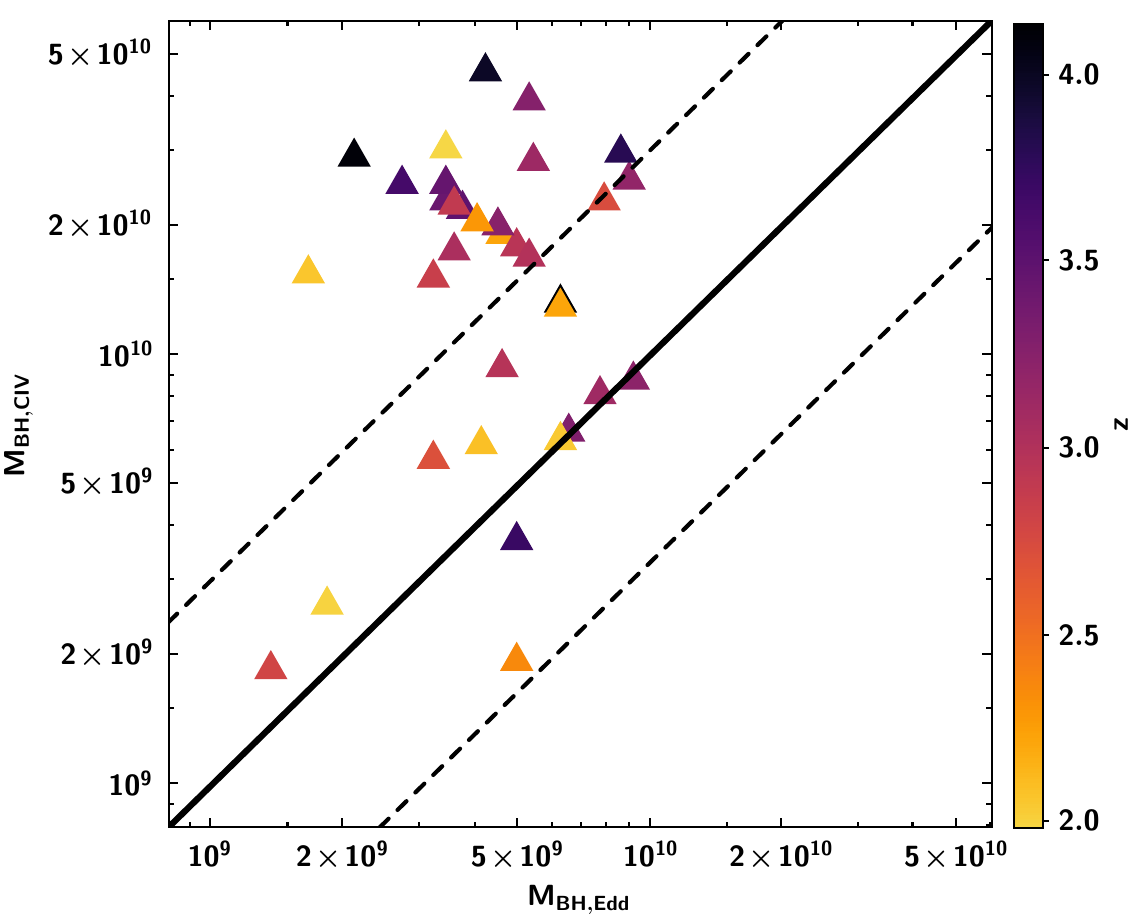}
  \hfill
  \includegraphics[width=0.32\linewidth]{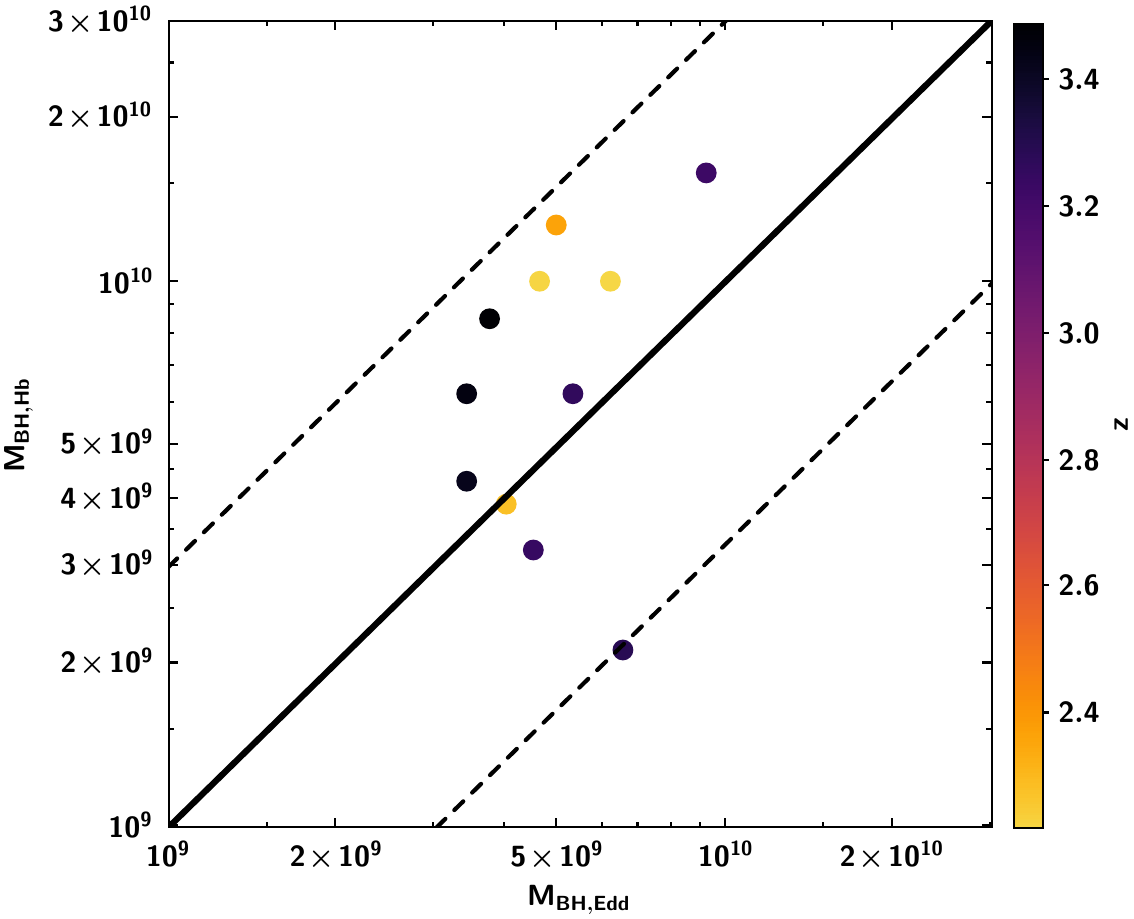}
  \caption{\textbf{Left panel:} \mbh\ obtained with the X-ray scaling method plotted vs. the values corresponding to sources accreting at the Eddington level. \textbf{Middle panel:} \mbh\ obtained with the SE method using the \ion{C}{iv} line  plotted vs. the Eddington values. \textbf{Right panel:} \mbh\ obtained with the SE method using the H$\beta$ line  plotted vs. the Eddington values. The symbols are color coded based on redshift. The continuous black line represents the one-to-one correlation and the dashed lines indicate departures by a factor of 3.}
  \label{fig:MEdd}
\end{figure*}


\section{Results}
\label{sec:Results}
\subsection{Black hole masses in the WISSH sample}
\label{sec:WISSH}
To assess the validity of the SE method, we utilize the X-ray scaling method, which in principle is valid for any moderately to highly accreting BH system (we briefly summarize the main characteristics of this method in the Appendix). However, one should bear in mind that this method is based on a simple Comptonization model (BMC), which was developed to study nearby objects. Therefore, we must first test whether this X-ray-based method yields reasonable results also at high redshifts.

The X-WISSH sample offers the most direct way to test this hypothesis. By construction, this sample contains hyperluminous quasars, located at high redshift, with broadband SEDs strongly dominated by the AGN component \citep{Duras2020}. It is reasonable to assume that these sources are accreting around the Eddington level (substantially lower accretion rates of the order of 0.1 would imply unrealistically high values of \mbh\ of the order of $10^{11}~\mathrm{M}_\odot$ or more). Supporting this hypothesis are the considerably high values of the X-ray bolometric correction, $K_\textrm{X}=L_\textrm{bol}/L_\textrm{Edd}$, which, for the 12 sources analyzed here, ranges from 39 to 1066 with an average of 273, implying that these sources are X-ray weak, as expected for highly accreting objects. Additionally, Fig.~9 of \citet{Duras2020} shows the X-WISSH sample clusters around $\log(\lambda_\textrm{Edd})=0$ in the $K_\textrm{X} - \log(\lambda_\textrm{Edd})$ plot.
With this simple assumption, $L_\mathrm{bol}=L_\mathrm{accr}=L_\mathrm{Edd}$, we can derive reasonable estimates for the BH masses by using $M_{\rm BH,Edd}=L_\mathrm{bol}/(1.3\times10^{38}~{\rm erg~s^{-1}})$, which can then be compared with the values obtained with the X-ray scaling method.

The left panel of Fig.~\ref{fig:MEdd} shows that the masses computed with the X-ray scaling method $M_{\rm BH,X}$ (plotted along the y-axis) are in general a good agreement with the $M_{\rm BH,Edd}$ values. We quantified this apparent agreement using $\langle {\rm max}(M_{\rm BH}\, {\rm ratio}) \rangle$, which is the mean of the maximum value between $M_{\rm BH,X}/M_{\rm BH,Edd}$ and $M_{\rm BH,Edd}/M_{\rm BH,X}$, and obtained 2.61 (with $\sigma/\sqrt{n}=0.93$, where $n$ is the number of objects). We then iteratively decreased $M_{\rm BH,X}$ using progressively smaller multiplicative factors to find a better agreement with $M_{\rm BH,Edd}$ and found a marginal improvement ($\langle {\rm max}(M_{\rm BH}\, {\rm ratio}) \rangle=2.33$, $\sigma/\sqrt{n}=0.62$) when using a multiplicative factor of 0.65, that is, when $M_{\rm BH,X}$ is decreased by 35\%, which is within the typical uncertainties of this method. 

When the same test is carried out on \mbh\ values obtained from the SE method using the \ion{C}{iv} line (illustrated in the middle panel of Fig.~\ref{fig:MEdd}), it is evident that a sizable portion of the values is overestimated, confirming that these measurements have a tendency to yield unreliable estimates in highly accreting objects. Using the same diagnostic tool for the comparison between the values computed with the \ion{C}{iv} line and the Eddington level \mbh\ (35 objects have both measurements), we get $\langle {\rm max}(M_{\rm BH}\, {\rm ratio}) \rangle = 4.36$. 

On the other hand, the \mbh\ estimates derived from the SE method using the H$\beta$ line appear to be broadly consistent with the $M_{\rm BH,Edd}$ values (right panel of Fig.~\ref{fig:MEdd}). In this case (with only 11 objects), we get $\langle {\rm max}(M_{\rm BH}\, {\rm ratio}) \rangle = 1.82$. Unfortunately, only three sources have \mbh\ estimated with both the X-ray-based method and the H$\beta$-based SE method; therefore, it is not possible to assess the consistency between these two methods using this limited sample. 

Since in this sample, there are only 12 and 11 objects with \mbh\ determined with the X-ray method and with the SE method using the H$\beta$ line, respectively, a statistical comparison with the corresponding Eddington limit values yields inconclusive results. Both sets of \mbh\ values appear to be consistent with the Eddington ones according to a Kolmogorov-Smirnov test ($P_\mathrm{KS}=0.19$ and 0.37, respectively), and their Spearman correlation coefficients indicate a positive but not statistically significant correlation with $r=0.40$ ($P_\mathrm{S}=0.19)$ and $r=0.29$ ($P_\mathrm{S}=0.38)$, respectively. On the other hand, the 35 sources with \mbh\ determined via the \ion{C}{iv} line make it possible to carry out a meaningful statistical comparison, which demonstrates that the distribution of these masses is not consistent with the Eddington limit one at high significance level ($P_\mathrm{KS}=7\times10^{-7}$) and that there is a weak but not statistically significant positive correlation ($r=0.10$, $P_\mathrm{S}=0.54)$.

In summary, exploiting the unique properties of the X-WISSH sample (highly accreting hyperluminous quasars with well-constrained bolometric luminosity and X-ray coverage), we can conclude that \ion{C}{iv}-based SE measurements are an unreliable estimator of \mbh, whereas both the X-ray scaling method and the H$\beta$-based SE method appear to yield reasonable results. Nevertheless, a larger sample of X-ray-bright, highly accreting AGN is necessary to carry out a quantitative comparison between these two indirect methods.

\begin{figure*}
  \includegraphics[width=0.32\linewidth]{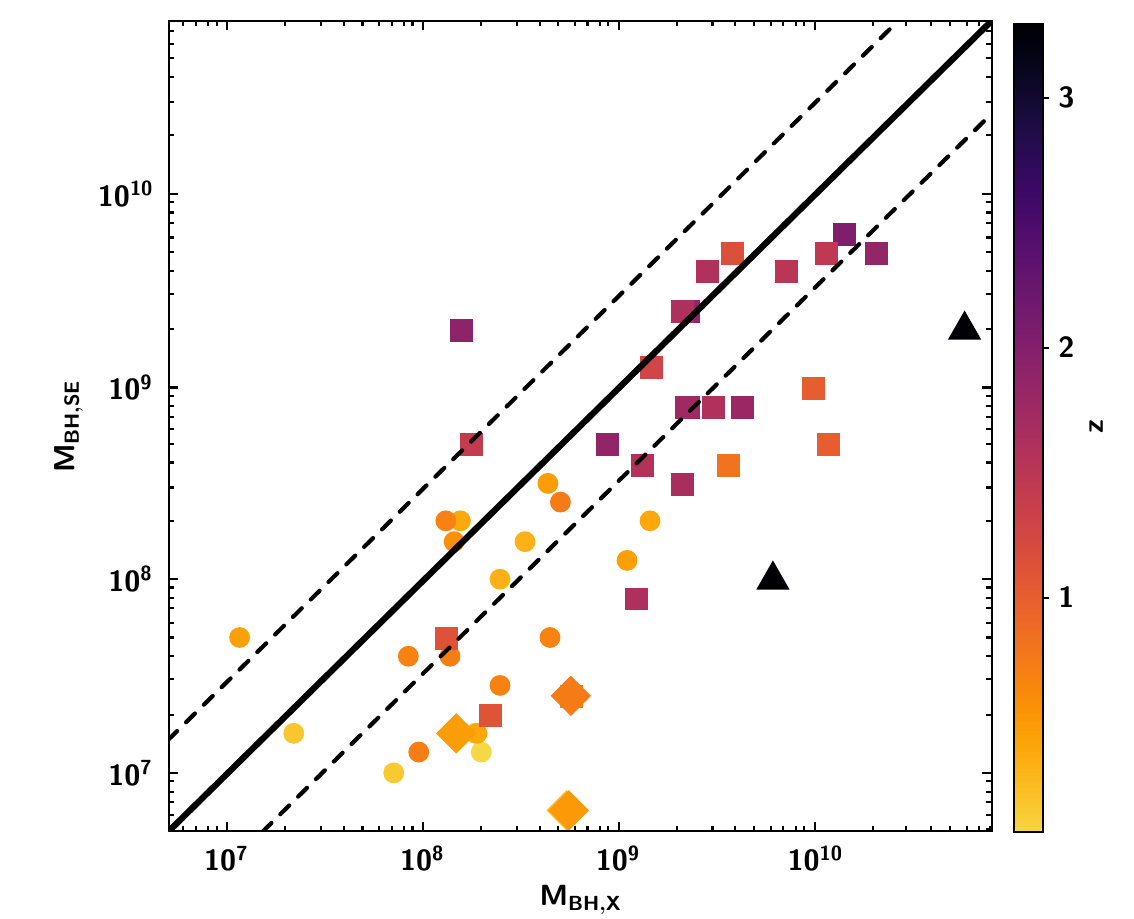}
  \hfill
  \includegraphics[width=0.32\linewidth]{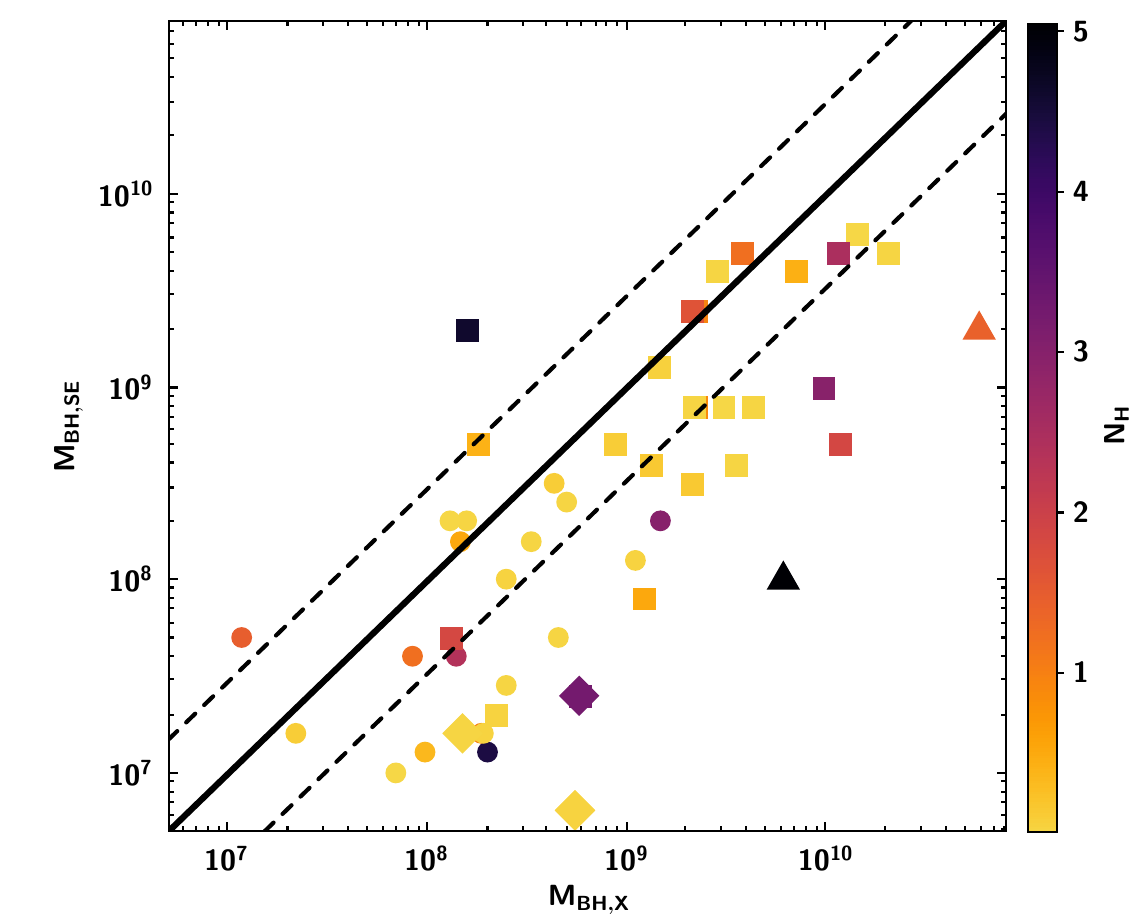}
  \hfill
  \includegraphics[width=0.32\linewidth]{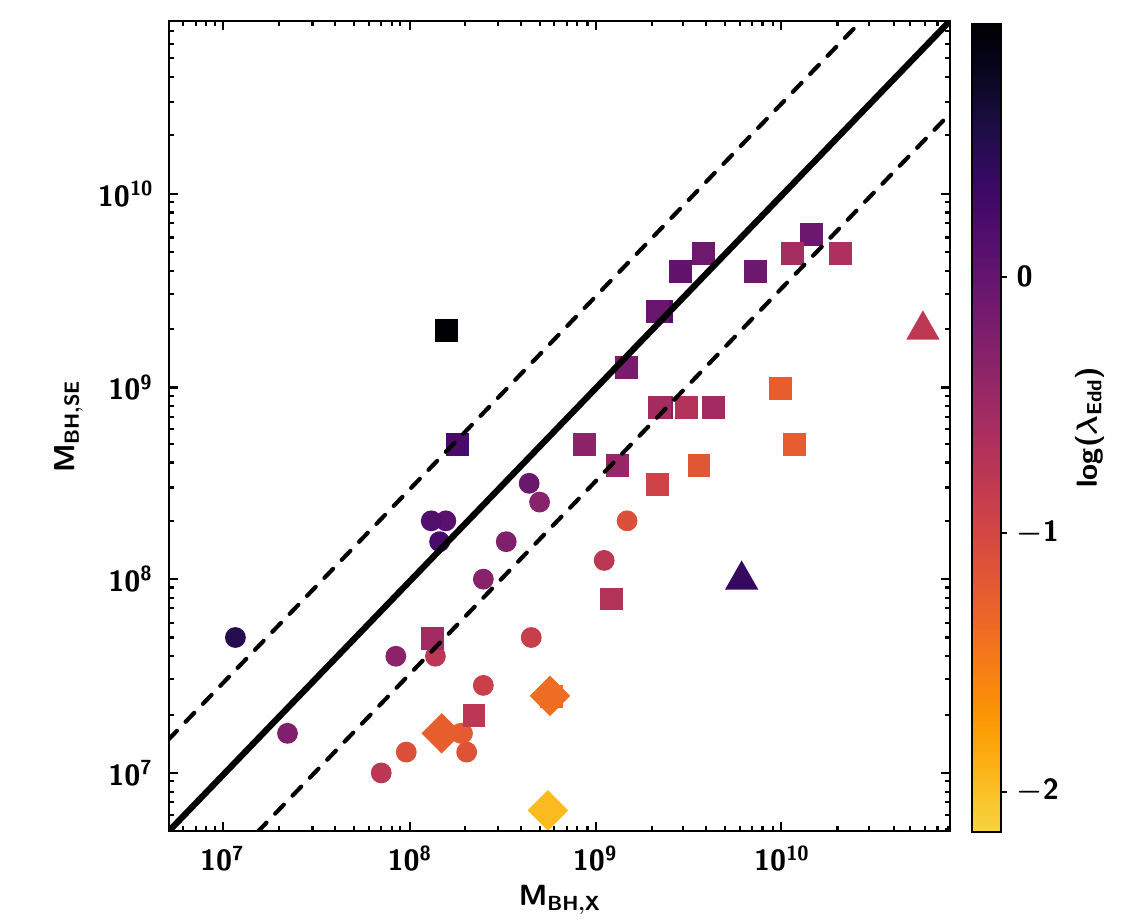}
  \caption{\textbf{Left panel:} \mbh\ obtained from the SE method using the H${\rm \beta}$ line plotted vs. the values derived with the X-ray scaling method, with color-coded symbols indicating the redshift of the sources.   \textbf{Middle panel:} same as the left panel with color-coded symbols illustrating the intrinsic absorption \nh\ in units of $10^{22}~{\rm cm^{-2}}$. \textbf{Right panel:} same as the left panel with color-coded symbols describing the level of accretion rate, as defined by $\log(\lambda_{\rm Edd})$. Circles indicate \mbh\ based on the H$\beta$ line, squares are values based on the \ion{Mg}{ii} line, whereas the two triangles indicate the values based on the \ion{C}{iv} line; the diamonds represent the sources whose \mbh\ values were flagged as unreliable in \citet{Rakshit2020}.}
  \label{fig:MxMse}
\end{figure*}

\begin{figure}
  \includegraphics[width=\columnwidth]{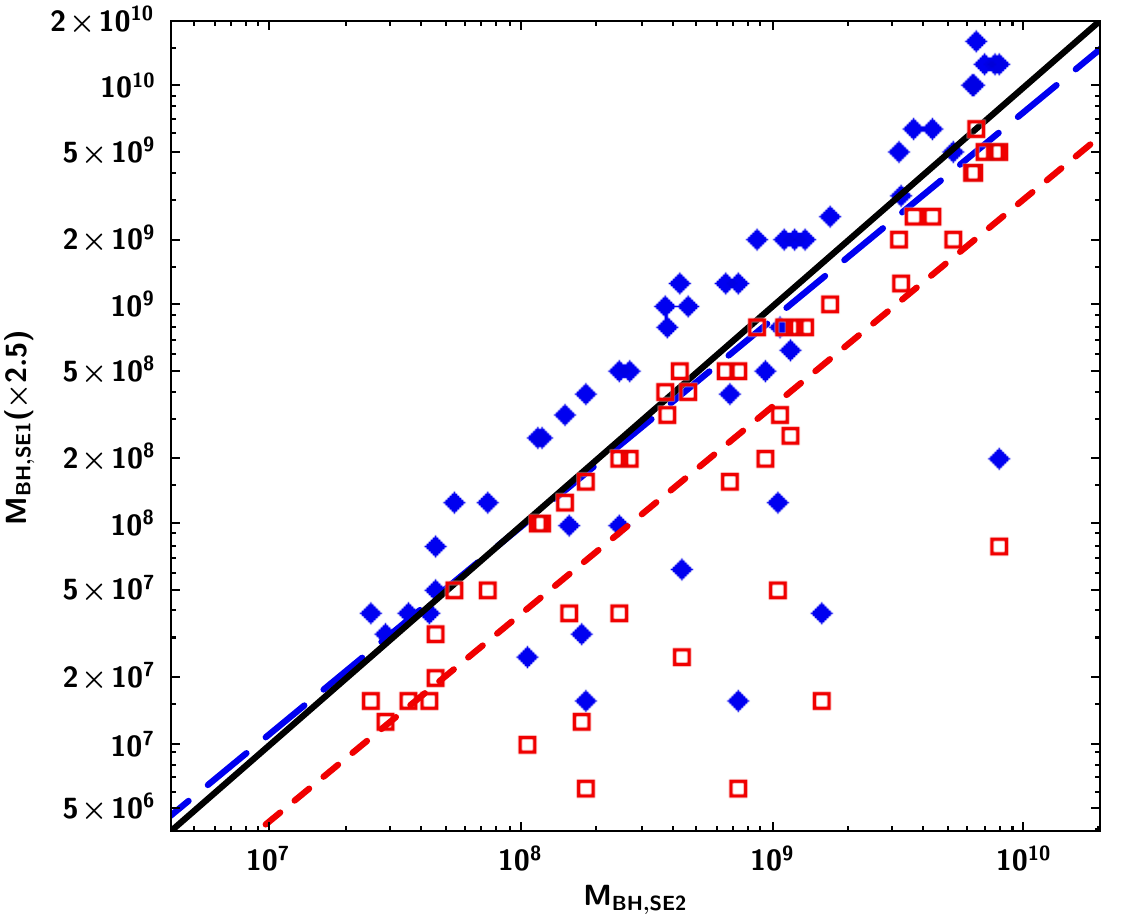}
  \caption{\mbh\ obtained from the SE method from \citet{Rakshit2020} plotted vs. the corresponding values from \citet{Wu2022}. The open squares are the unshifted values and the red short-dashed line indicates the best linear fit. The blue diamonds represent the \citet{Rakshit2020} shifted by a factor 2.5; in this case the blue long-dashed line, which represents the best linear fit, nearly overlaps the one-to-one correlation illustrated by the continuous black line.}
  \label{fig:Mse1Mse2}
\end{figure}

\subsection{Black hole masses in the X-HESS sample}
\label{sec:XHESS} 
The X-HESS sample allows a direct comparison between the \mbh\ estimates obtained with the X-ray scaling method and those derived with the SE method based mostly on the H$\beta$ and \ion{Mg}{ii} lines. Of the original 60 AGN presented by \citet{Laurenti2024}, 50 objects have adequate count rates and $\Gamma$ in the proper range to apply the X-ray scaling method. A visual comparison between these two indirect methods is illustrated in Fig.~\ref{fig:MxMse}, where the \mbh\ values obtained from the SE method using different lines depending on the source redshift (specifically, 23 are based on the H$\beta$ line and represented by circles, 25 on the \ion{Mg}{ii} line and illustrated with squares and two on the \ion{C}{iv} line and described by triangles) are plotted vs. the values derived with the X-ray scaling method (the diamonds represent the four sources whose \mbh\ values were flagged as unreliable in \citet{Rakshit2020}).
The figure has three panels with different color schemes illustrating, respectively:  the redshift $z$ (left panel), the intrinsic absorption measured by \nh\ in units of $10^{22}~{\rm cm^{-2}}$\ (middle panel), and  the accretion rate level defined by $\log(\lambda_{\rm Edd})$ (right panel).

A visual inspection of Fig.~\ref{fig:MxMse} clearly indicates the existence of a strong positive correlation between the \mbh\ values obtained with these two indirect methods. This is quantitatively confirmed by a Spearman's rank correlation analysis that yields a coefficient $r=0.74$ and relative probability $P_{\rm S} \simeq 10^{-9}$. The same figure also indicates the presence of an offset between the SE \mbh\ values and the estimates derived with the X-ray scaling method, with the former values that appear to be systematically lower than the latter ones. This conclusion is confirmed by a linear regression carried out with the \textsc{linmix\textunderscore err} routine, which accounts for both errors on the x-axis (we used the percent errors directly derived by the X-ray scaling method, which typically range between 25\% and 50\%) and on the y-axis (for the SE values we have assumed a typical uncertainty of 0.4 dex), and yields $M_{\mathrm{BH,SE}}= (0.64 \pm 0.15) + (0.86\pm0.02)\times M_{\mathrm{BH,X}}$  with an RMS deviation of 0.62. We note that the same linear regression routine will be used throughout the paper.

The different color schemes used in the three panels of Fig.~\ref{fig:MxMse} may help shed some light on the offset between these two indirect methods. Based on the left panel, we can rule out that the distance (parameterized by the redshift) is a major factor: with the exception of the two \mbh\ values based on the \ion{C}{iv} line, most of the values that fall below the one-to-one correlation are located at relatively low redshift, where both indirect methods are well calibrated.  On the other hand, the middle panel suggests that some of the largest discrepancies are either associated  with high intrinsic absorption or with \mbh\ values that were flagged as unreliable in \citet{Rakshit2020}. 
Finally, the right panel of Fig. \ref{fig:MxMse} suggests that only the most super-Eddington source is overestimated by the SE method, whereas there appears to be no discrepancy between these two indirect methods for high and moderately high accreting AGN. 

This result appears puzzling for two reasons. Firstly, the standard SE method is known to underestimate \mbh\ in highly accreting AGN \citep{Du2015,Martinez2019}; therefore, one would expect that the discrepancy between these two methods would occur for the most highly accreting objects. Secondly, both the H$\beta$-based and \ion{Mg}{ii}-based SE method and the X-ray scaling method are fully consistent with \mbh\ values of the moderately accreting AGN, derived with the reverberation method in the local universe \citep{Bentz2013, Gliozzi2011}. 

\begin{figure}
  \hspace{1pt}
  \includegraphics[width=0.93\columnwidth]{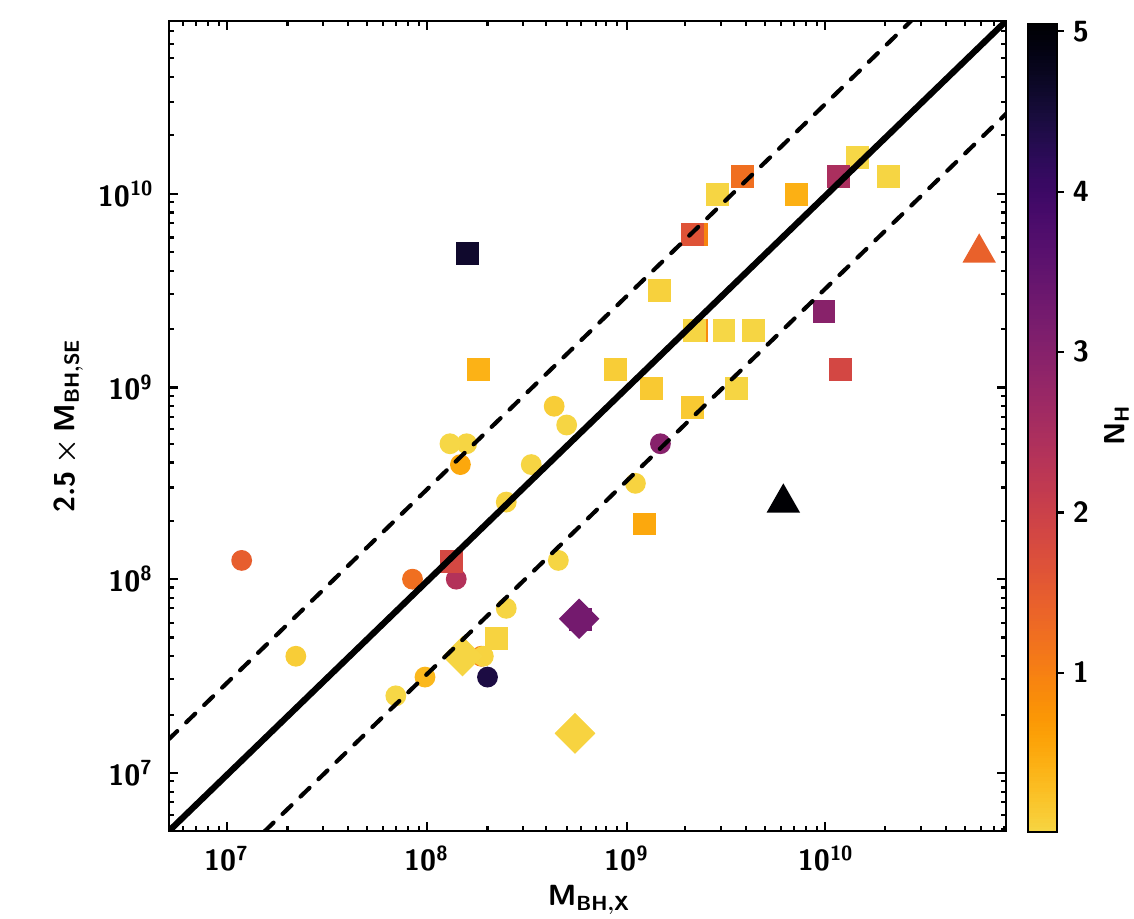}

  \includegraphics[width=\columnwidth]{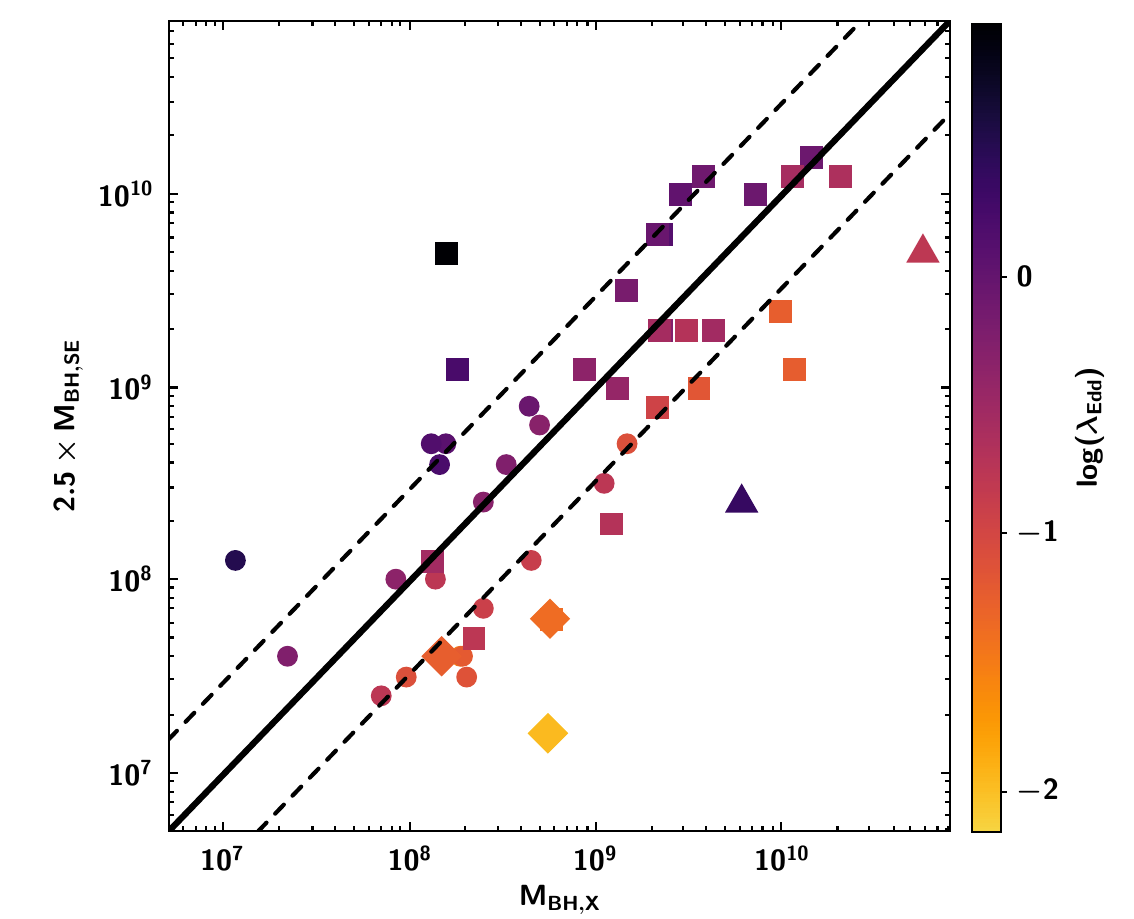}
  \caption{\textbf{Top panel:} SE \mbh\ values multiplied by a factor of 2.5 plotted vs. the X-ray-based values, with color-coded symbols indicating the intrinsic absorption \nh\ in units of $10^{22}~{\rm cm^{-2}}$. \textbf{Bottom panel:} same as the top panel, with color-coded symbols indicating $\log(\lambda_{\rm Edd})$. The symbols are the same as the ones used in Fig. \ref{fig:MxMse}.}
  \label{fig:MxMse-shift}
\end{figure}

To ensure that moderately accreting AGN yield consistent values, \mbh\ should be recalibrated either by decreasing the values obtained with the X-ray method or by increasing those obtained with the SE method (or by a combination of both). Although in principle both corrections are equally plausible based on the results in Section \ref{sec:WISSH}, one must keep in mind that H$\beta$-based SE method values used in that section were based on high-quality data obtained with dedicated campaigns \citep{Vietri2018}. On the other hand, the H$\beta$-based and \ion{Mg}{ii}-based \mbh\ values used in \citet{Laurenti2024} are taken from the catalog of spectral properties of quasars from the Sloan Digital Sky Survey Data Release 14 of \citet{Rakshit2020} and computed from an automated spectral decomposition analysis using limited quality data. 

Importantly, as noted by \citet{Rakshit2020}, the average width of the H$\beta$ line is systematically smaller by 0.111 dex (with a dispersion of 0.140 dex) compared to other catalogs based on the same data set \citep{Calderone2017}. This suggests that the \mbh\ values based on \citet{Rakshit2020} can be systematically underestimated, since the \mbh\ in the SE method has a quadratic dependence on the line width. To test this hypothesis, we made a direct comparison of the \citet{Rakshit2020} values with the \mbh\ from the catalog of \citet{Wu2022}. The results are illustrated in Fig.~\ref{fig:Mse1Mse2}, where the black continuous line represents the one-to-one correlation, the red open squares are the unshifted values, and the blue diamonds indicate the \citet{Rakshit2020} values increased by a factor of 2.5. The fact that the linear best fit of the shifted values (blue long-dashed line) nearly overlaps the one-to-one correlation (as opposed to the red short-dashed line indicating the best linear fit of the unshifted data, which lies well below) confirms that the \citet{Rakshit2020} \mbh\ values are systematically underestimated.

Indeed, when we increase the SE \mbh\ by a factor of 2.5 and compare them to the X-ray scaling values, the bulk of the AGN becomes consistent (see Fig.~\ref{fig:MxMse-shift}). We note that very similar figures are obtained (but not shown here) when the \mbh\ values from the catalog of \citet{Wu2022} without any correction factor are plotted versus the X-ray scaling values.

 A close look at Fig.~\ref{fig:MxMse-shift}, where the recalibrated SE \mbh\ values are plotted vs. the X-ray based estimates, reveals that 1) the vast majority of the \mbh\ estimates are now in full agreement; 2) the few objects that still have significantly lower SE \mbh\ values compared to the corresponding X-ray ones are either based on the \ion{C}{iv} line (triangular symbols) or are substantially absorbed (see the top panel of the figure, where the color-coded scheme describes the intrinsic absorption) or were flagged in the \citet{Rakshit2020} catalog, which are represented by diamond symbols (there are actually four flagged sources, but two sources perfectly overlap in these plots); 3) the few SE \mbh\ values that are substantially overestimated with respect to the X-ray estimates are sources accreting at super-Eddington rates (see the bottom panel of the same figure, where the color-coded scheme describes the accretion rate in Eddington units). 

In summary, using the X-HESS sample of \citet{Laurenti2024}, we have demonstrated that there is a strong positive correlation between the values of \mbh\ obtained with the SE method and those from the X-ray scaling method, indicating that both methods provide the same relative \mbh\ values. However, our analysis also reveals that the SE-based measurements from the \citet{Rakshit2020} catalog are  systematically underestimated by a factor of 2.5. Once the SE-based values of that specific catalog are recalibrated to ensure that they are consistent for moderately accreting AGN, then, in agreement with the current understanding of the limitations of the standard SE method, the discrepancies are limited to extremely highly accreting objects, whose \mbh\ is significantly overestimated by the standard SE method, and to substantially absorbed objects, whose \mbh\ is instead underestimated by the SE method.

\subsection{Implications of using different indirect methods}
\label{sec:Implications}
\begin{figure} 
  \includegraphics[width=\columnwidth]{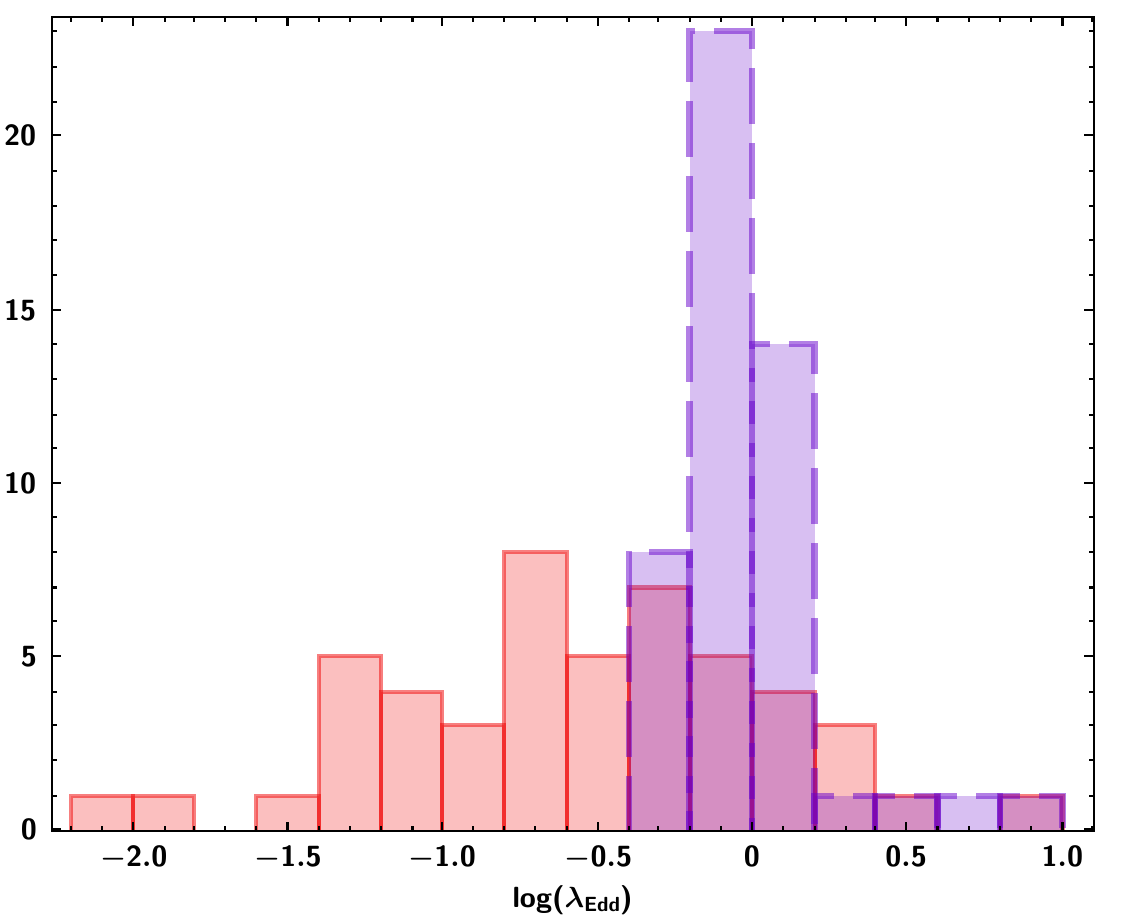}
  \caption{Histograms of the $\log(\lambda_{\rm Edd})$ distributions obtained with the X-ray scaling method (red) and the SE method (purple).}
  \label{fig:histloglaEd}
\end{figure}

Although the X-ray scaling method and the SE method are broadly consistent with each other, the use of \mbh\ values obtained with the former method specifically from the catalog of \citet{Rakshit2020} (which we now know to systematically underestimate the \mbh), including a few sources with mass based on \ion{C}{iv} (which our analysis in Section \ref{sec:WISSH} confirms to be an unreliable estimator of virial mass) and a few sources that were flagged and therefore deemed unreliable, can have important implications and lead to markedly different conclusions. To illustrate this point, we compare different results obtained using \mbh\ estimates from the X-ray scaling method with some of the corresponding findings derived by \citet{Laurenti2024} with a thorough analysis utilizing the \mbh\ values from the catalog of \citet{Rakshit2020}.

\noindent{\bf Eddington ratio distribution.} The first substantial difference is illustrated in Fig.~\ref{fig:histloglaEd}, which shows the distributions of the accretion rate values. The distribution obtained with the X-ray method (red color) spreads between $\log(\lambda_{\rm Edd})=-2$ and 1, whereas by construction the distribution of the X-HESS of \citet{Laurenti2024} based on SE values from \citet{Rakshit2020} is solely restricted to highly accreting objects. The visual difference is quantitatively confirmed by a Kolmogorov--Smirnov test, which yields  $K=0.74$ and associated probability $P_{\rm K} \simeq 10^{-11}$.

\begin{figure} 
  \includegraphics[width=\columnwidth]{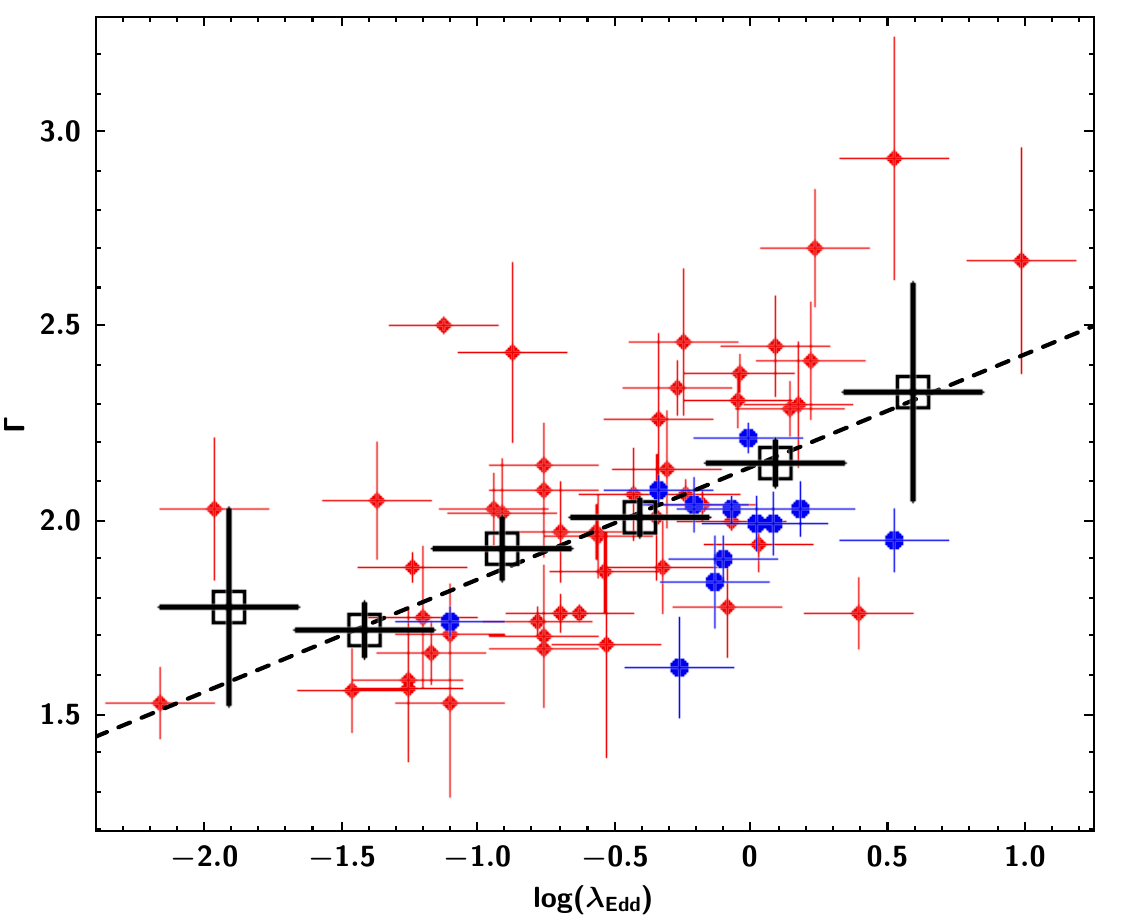}
  \caption{X-ray photon index $\Gamma$ plotted vs. $\log(\lambda_{\rm Edd})$. The dashed line represents the best fit including both the X-HESS sample (red diamonds) and WISSH sample (blue circles), 
  $\Gamma=(2.14\pm0.01)+(0.29\pm0.01)\times \log(\lambda_{\rm Edd})$. The black squares represent the arithmetic mean obtained by binning the data.}
  \label{fig:GaloglaEd}
\end{figure}
\noindent{\bf Photon index vs. accretion rate.} When the photon index $\Gamma$ is plotted vs. $\log(\lambda_{\rm Edd})$, a strong positive correlation is obtained with Spearman coefficient $r=0.54$ and relative probability $P_{\rm S} =8\times 10^{-6}$ and a best fit of 
$\Gamma=(2.14\pm0.01)+(0.29\pm0.01)\times \log(\lambda_{\rm Edd})$ (RMS = 0.26) obtained considering both the X-HESS data (represented by the red diamonds in Fig.~\ref{fig:GaloglaEd}) and WISSH data (blue circular symbols). Note that the same strong correlation with a slightly steeper positive slope, $\Gamma=(2.19\pm0.01)+(0.32\pm0.01)\times \log(\lambda_{\rm Edd})$, is derived when only the X-HESS sample is used, at odds with the findings of \citet{Laurenti2024}, who did not find any significant correlation using the same sample.

\begin{figure}
  \includegraphics[width=\columnwidth]{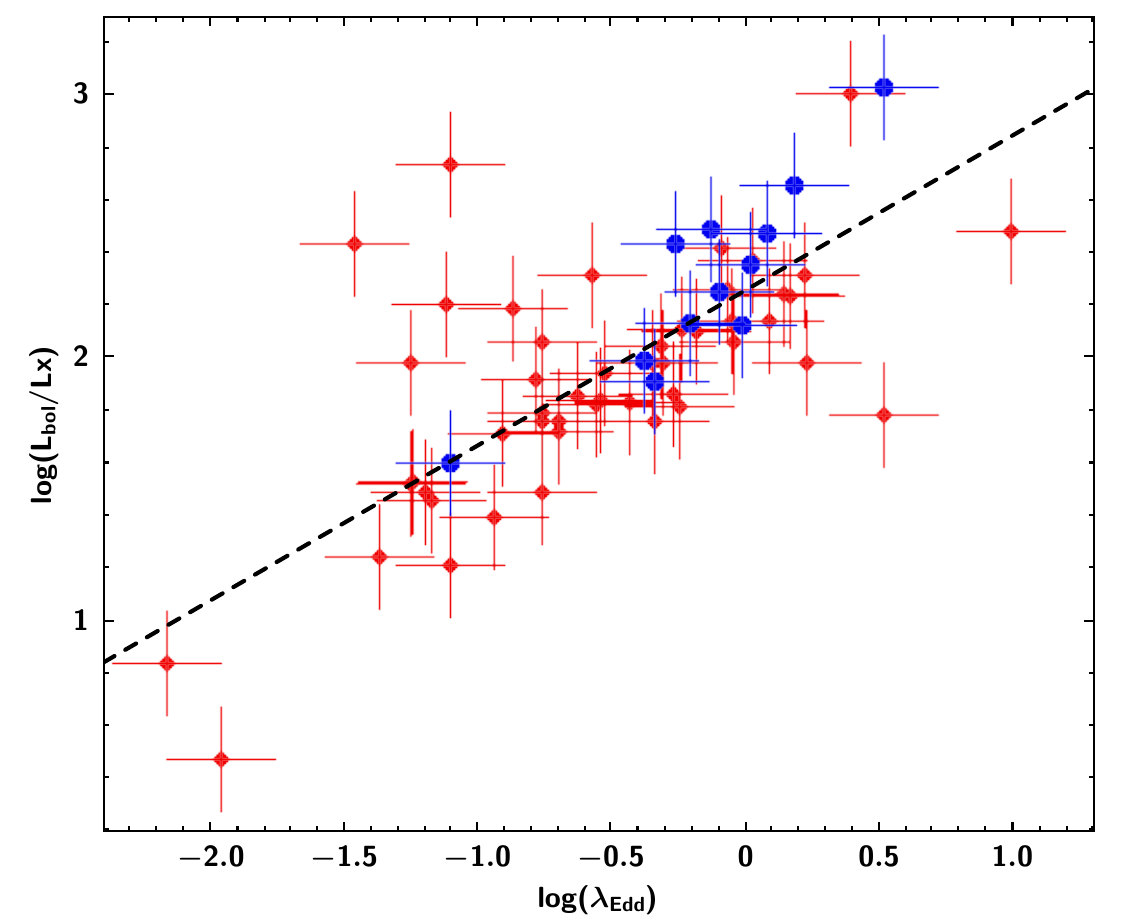}

  \includegraphics[width=\columnwidth]{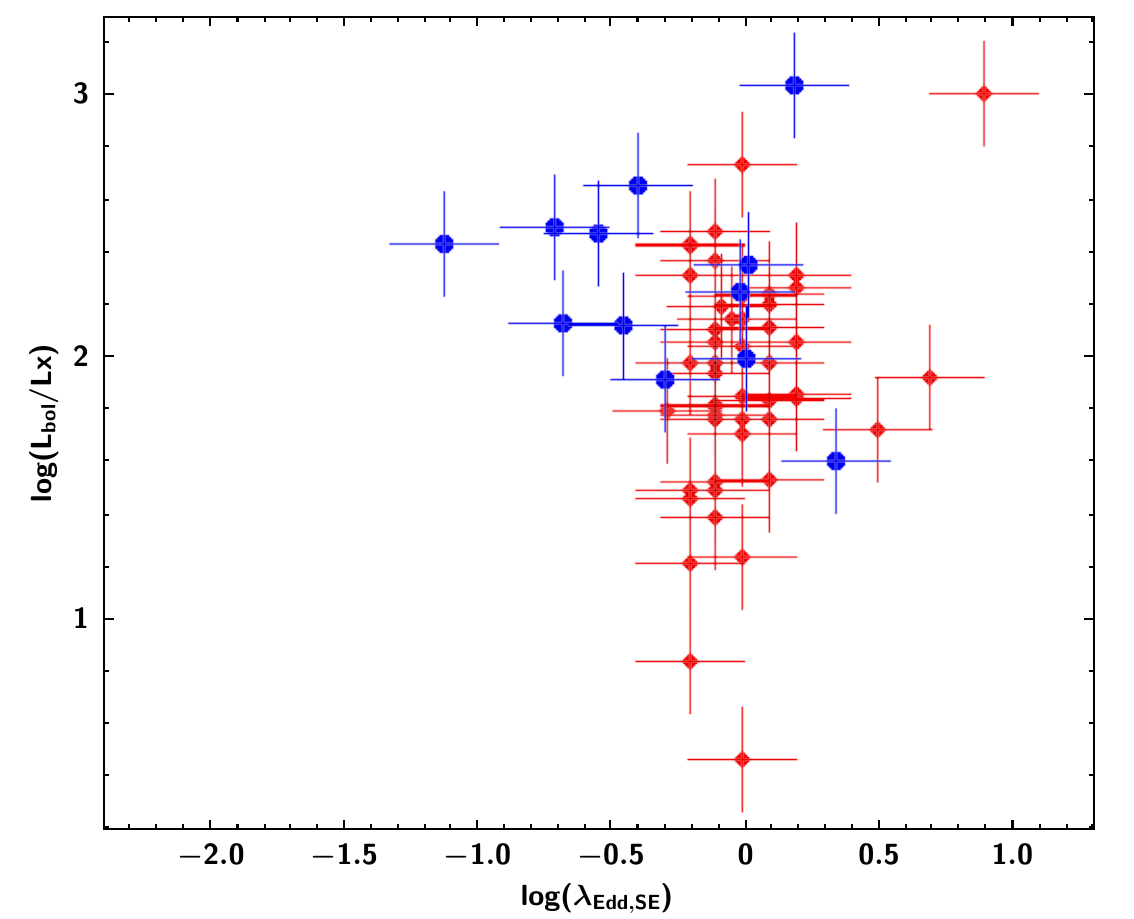}
  \caption{\textbf{Top panel:} Logarithm of the X-ray bolometric correction factor $\log(L_\mathrm{bol}/L_\mathrm{X})$  plotted vs. $\log(\lambda_\mathrm{Edd})$, where the latter is computed using the X-ray scaling method. The dashed line represents the best fit including both the X-HESS sample (red diamonds) and WISSH sample (blue circles), $K_\mathrm{X}=(2.26\pm0.01)+(0.59\pm0.01)\times \log(\lambda_\mathrm{Edd})$. \textbf{Bottom panel:} Same plot with $\log(\lambda_\mathrm{Edd})$ derived using SE measurements. No significant correlation is obtained.}
  \label{fig:logLboLxloglaEd}
\end{figure}
\noindent{\bf X-ray bolometric correction vs. accretion rate.} Markedly different results are also obtained when $\log(L_{\rm bol}/L_{\rm X})$ (the logarithm of the X-ray bolometric correction factor $K_{\rm X}$) is plotted  vs. $\log(\lambda_{\rm Edd})$, as illustrated in the top panel of Fig.~\ref{fig:logLboLxloglaEd}, where we used the \mbh\ measured with the X-ray method to compute the Eddington ratio. A strong positive correlation is obtained with the X-ray-based values, as confirmed by a Spearman coefficient $r=0.65$ and relative probability $P_{\rm S} =1.9\times 10^{-8}$ and a best fit of  $K_{\rm X}=(2.26\pm0.01)+(0.59\pm0.01)\times \log(\lambda_{\rm Edd})$ (RMS~=~0.33). Conversely, no correlation at all is present when SE data are used, as illustrated in the bottom panel of Fig. \ref{fig:logLboLxloglaEd}, and confirmed by a statistical analysis: $r=-0.04$ ($P_{\rm S} =0.75$), $K_{\rm X}=(1.98\pm0.01)-(0.03\pm0.04)\times \log(\lambda_{\rm Edd})$.

\noindent{\bf Soft excess.} Finally, we focus on the soft excess. To this end, we extend our spectral analysis to the 0.3--10 keV range (in the observer frame) and add a blackbody component to our baseline model to fit the spectra of sources that show a clear excess (19 in our original sample of 50 AGN), when extrapolating the Comptonized model used to fit the hard X-ray spectrum. 

First of all, we verify that different measurements of the strength of the soft excess provide similar results. Indeed, a Spearman analysis ($r=0.89$ and relative probability $P_{\rm S} =2.4\times 10^{-7}$) indicates that $\log(\mathit{SX1})=\log(L_{\rm bb}/L_{\rm bmc})_{\rm 0.5-2 keV}$, the quantity similar to that used by \citet{Laurenti2024}, is strongly correlated with $\log(\mathit{SX3})=\log(L_{\rm bb, 0.5-2\,keV}/L_{\rm Edd})$, the quantity introduced in \citet{Gliozzi2020} and that will be used again here (for both measurements of the soft excess strength, the uncertainties were calculated with error propagation). This is confirmed by linear regression analysis that yields a best fit of $\log(\mathit{SX1})=(1.58\pm0.04)+(0.63\pm0.02)\times \log(\mathit{SX3})$ (RMS = 0.32).

\begin{figure} 
  \includegraphics[width=\columnwidth]{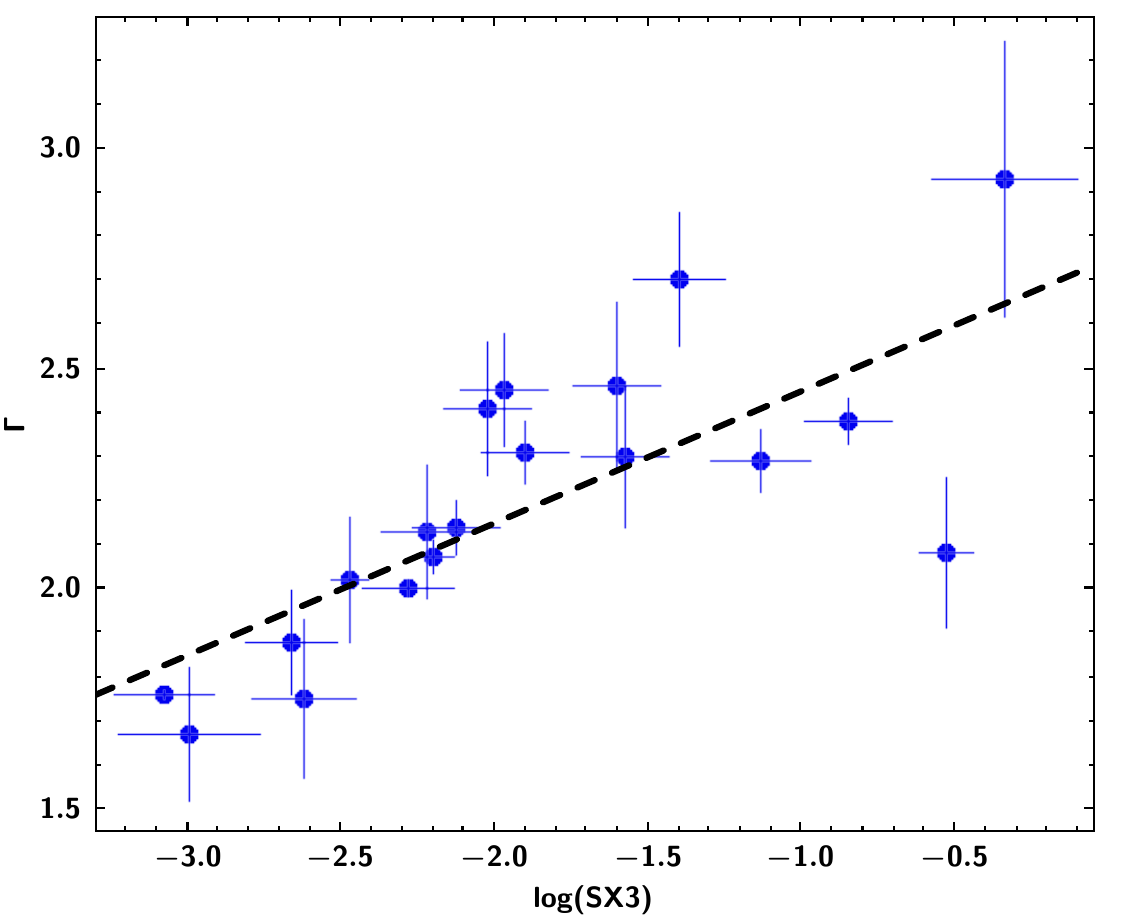} 
  \caption{Photon index plotted vs. the soft excess strength $\log(\mathit{SX3})$ with the best-fit correlation, $\Gamma=(2.75\pm0.01)+(0.30\pm0.01)\times \log(\mathit{SX3})$.}
  \label{fig:logSX3}
\end{figure}

Fig.~\ref{fig:logSX3} reveals the presence of a strong positive correlation between $\Gamma$ and $\log(\mathit{SX3})$, described by the best-fit correlation, $\Gamma=(2.75\pm0.01)+(0.30\pm0.01)\times \log(\mathit{SX3})$ (RMS = 0.29), and further confirmed by a Spearman coefficient of $r=0.79$ and relative probability $P_{\rm S} =4.5\times 10^{-5}$. Note that similar conclusions at lower significance level are obtained also using $\log(\mathit{SX1})$: $\Gamma=(2.08\pm0.02)+(0.29\pm0.03)\times \log(\mathit{SX1})$, $r=0.58$ ($P_{\rm S} =9.6\times 10^{-3}$). This is  again at odds with the results of \citet{Laurenti2024}, who inferred the presence of a negative correlation between the photon index and the soft excess strength.

In our statistical analysis we also found a positive correlation between $\log(\mathit{SX3})$ and $\log(\lambda_{\rm Edd})$: 
$\log(\mathit{SX3})=(-1.62\pm0.05)+(1.07\pm0.11)\times \log(\lambda_{\rm Edd})$, $r=0.63$ ($P_{\rm S} =3.6\times 10^{-3}$) (RMS = 0.67), whereas no significant correlation was obtained when the same parameter was plotted versus the hard X-ray luminosity: $\log(\mathit{SX3})=(2.67\pm2.85)-(0.10\pm0.06)\times \log(L_{\rm X})$, $r=-0.12$ ($P_{\rm S} =0.63$).

\section{Discussion}
The primary goal of this work is to assess the reliability of the SE method applied to distant AGN. This task is extremely challenging, because direct dynamical methods cannot be applied to distant objects, and other commonly used indirect methods, such as those based on various relationships between \mbh\ and galaxy bulge properties observed in lower-luminosity low-accreting objects in the local universe, may not be suited for highly accreting distant objects. Moreover, since the final goal is to study SMBH-galaxy coevolution over cosmic time, one cannot use the \mbh\ inferred from these correlations.

To assess the reliability of the SE method in distant AGN we used the X-ray scaling method, which has proven to be one of the most reliable indirect methods based on our recent study of a volume-limited, hard-X-ray-selected sample of AGN with \mbh\ dynamically determined \citep{Gliozzi2024}.

Since this method is based on a simple Comptonization model (BMC) developed for nearby sources and was essentially used only for local AGN, we first ran a sanity check by applying it to a sample of distant luminous quasars (the X-WISSH sample) with well-defined multiwavelength SEDs completely dominated by the AGN emission \citep{Duras2020} and for which it is reasonable to assume that accretion occurs at rates close to the Eddington value. This test indicated that the X-ray scaling method yields reasonable \mbh\ values also for distant quasars. Similarly does the SE method based on the H$\beta$ line, whereas the SE method based on the \ion{C}{iv} line shows a tendency to systematically overestimate \mbh, confirming the conclusions of several studies that questioned the use of high-ionization lines, which are likely to be affected by outflows (e.g., \citealt{Denney2012, Denney2016}). Unfortunately, only three sources have \mbh\ measurements obtained with both the X-ray scaling method and the H$\beta$-based SE method, hampering a quantitative comparison between these two techniques using the X-WISSH sample.

The X-HESS sample presented by \citet{Laurenti2024}, which comprises X-ray-bright AGN that are  supposedly highly accreting and possess \mbh\ determined with the SE method using primarily the H$\beta$ or \ion{Mg}{ii} lines, makes it possible to carry out a systematic comparison between these two indirect methods. Fig.~\ref{fig:MxMse} and the associated analysis reveal a strong linear correlation between the two methods, which indicates that, despite the completely different assumptions, these two methods give consistent results. The same analysis however also shows that the bulk of the moderately accreting objects have systematic differences with the X-ray-based \mbh\ values being substantially larger than the corresponding SE estimates. 

This discrepancy, which is surprising because in the moderately accreting regime both methods are fully consistent with the values of \mbh\ obtained with the RM technique, can be resolved either by decreasing the X-ray values or by increasing the SE ones by a factor of 2.5.  The fact that the average FWHM of lines used to estimate \mbh\ from \citet{Rakshit2020} is systematically smaller than that from \citet{Calderone2017}, who use the same data but a different procedure to define and fit the spectral lines suggests that the \mbh\ values in the \citet{Rakshit2020} catalog are underestimated. Indeed, this hypothesis is confirmed by direct comparison between the \mbh\ values of \citet{Rakshit2020} and the corresponding ones presented by \citet{Wu2022} in the most recent version of the SDSS catalog for AGN (see Fig.~\ref{fig:Mse1Mse2}).
We note that the few \mbh\ values in the X-HESS sample that were obtained with more refined H$\beta$ measurements \citep{Marziani2014} are fully consistent with the X-ray values, when the sources are not substantially absorbed.

After the two methods are recalibrated by multiplying the SE values by a factor of 2.5 (see Fig. \ref{fig:MxMse-shift}), which is of the order of the systematic uncertainty associated  with this method (see, e.g., \citealt{Pancoast2014}), then we obtain a better overall agreement between the two methods. Now the few significant discrepancies are related to 1) sources whose \mbh\ and $\lambda_\mathrm{Edd}$ values were flagged for poor quality by \citet{Rakshit2020} (namely, source 6 SDSS J172255.24+320307.5, source 14 SDSS J221715.18+002615.0, source 22 SDSS J233317.38-002303.4, and source 34 SDSS J084153.99+194303.1, which are represented by diamonds in Fig. \ref{fig:MxMse-shift}); 2) sources whose \mbh\ was computed using the \ion{C}{iv} line (source 36 SDSS J090033.50+421547.0 and source source 54 SDSS J132654.95-000530.1, which are represented by triangles in Fig. \ref{fig:MxMse-shift}); 3) substantially absorbed sources, whose \mbh\ values appear to be underestimated by the SE method in agreement with the findings of \citet{Mejia-Restrepo2022}; and 4) sources accreting well above the Eddington ratio, for which the SE method overestimates \mbh, as expected from the work of \citet{Du2015,Du2018}.

While the broad agreement observed between the two methods without applying any recalibration (that is, using the \mbh\ values provided by \citet{Rakshit2020}, as done by \citet{Laurenti2024})
may suggest that either method could be used interchangeably (after all, a calibration factor of 2.5 is of the order of the typical uncertainty associated with the SE method), our analysis demonstrates this is not the case. 

Firstly, the distribution of the accretion rates derived from the use of the X-ray-based \mbh\ is substantially broader than the one presented by \citet{Laurenti2024} based on the optically based \mbh\ values from the catalog of \citet{Rakshit2020}, suggesting that this sample is not strictly restricted to highly accreting objects (see Fig. \ref{fig:histloglaEd}). 

Secondly, we find a strong positive correlation when the photon index is plotted versus $\log(\lambda_\mathrm{Edd})$, in contrast with \citet{Laurenti2024}, who found a weak positive correlation only when their data were combined with additional samples spanning a broader range of accretion rates. Our results confirm and strengthen the conclusion that $\Gamma$ is a faithful indicator of the accretion state of BH systems, as regularly seen in stellar-mass BH systems in their spectral transitions \citep{Remillard2006} and found in several studies based on AGN samples of different redshifts and luminosities \citep{Shemmer2008,Risaliti2009,Brightman2013,Brightman2016,Serafinelli2017}, including long-term spectral variability investigations of individual AGN \citep{Sobolewska2009}.

Additionally, the X-ray bolometric correction $K_\mathrm{X}$ plotted versus $\log(\lambda_\mathrm{Edd})$ shows a strong positive correlation only if the Eddington luminosity uses X-ray measurements (see Fig. \ref{fig:logLboLxloglaEd}), in agreement with the findings of \citet{Lusso2012} and \citet{Duras2020}. The lack of a clear positive correlation when we use the SE values (see the bottom panel of Fig. \ref{fig:logLboLxloglaEd}) suggests that the indiscriminate use of \mbh\ values from large catalogs based on automated analysis of spectral data with limited quality may be problematic.

Finally, we find a strong positive correlation between $\Gamma$ and the strength of the soft excess, in agreement with the results obtained by \citet{Bianchi2009,Boissay2016,Gliozzi2020}. Similarly to the latter study, we also obtain a positive correlation between the strength of the soft excess and the accretion rate, and no correlation at all when the strength of the soft excess is plotted vs. the hard X-ray luminosity. All these findings can be naturally explained in the framework where the soft excess is dominated by a warm corona \citep{Done2012,Rozanska2015,Petrucci2018}. In contrast, \citet{Laurenti2024} inferred the presence of a negative correlation between the photon index and the soft excess strength, which favors the ionized reflection model as the main cause of the soft excess, yet there is no evidence for strong reflection components in any of the spectra of the X-HESS sample.

Before reaching our conclusions, we try to leverage our main results to shed some light on the enigmatic nature of LRDs. Using a sample comprising high-redshift AGN, we find that both the X-ray bolometric correction and the photon index are strongly correlated with the accretion rate measured by $\lambda_\mathrm{Edd}$. This confirms that the extreme X-ray weakness of LRGs may be naturally explained by AGN in the highly accreting regime, as proposed by recent theoretical works \citep{Lupi2024,Pacucci2024,Lambrides2024,Madau2024}: high values of the X-ray bolometric correction factor ($K_\mathrm{X} >10^3$) coupled with a steep photon index
($\Gamma \geq 2.5$) may make moderately luminous very distant AGN virtually undetectable by current X-ray observatories, since the already intrinsically weak X-ray emission is concentrated in the soft part of the spectrum, which falls below the lower energy threshold of current instruments by virtue of the high redshift of these sources. Additionally, our analysis suggests that very highly accreting AGN are likely to have values of \mbh\ substantially overestimated by the standard SE method, lessening some of the constraints on BH growing models and the extreme BH-to-galaxy mass ratios inferred for some of these sources.

\section{Conclusions}
To conclude, we summarize the main findings of this study.
   \begin{enumerate}
      \item We first verified that the X-ray scaling method yields reasonable estimates of \mbh\ also for distant AGN and quasars.
      \item We then used the X-ray scaling method to test the reliability of the SE method applied to highly accreting AGN. The strong correlation between the \mbh\ values obtained with these two different indirect methods is encouraging, given the very different assumptions of the two methods and suggests that the use of SE \mbh\ values for large population AGN studies is appropriate in most instances.
      \item Our comparison of the two methods, however, also indicates the presence of a relevant difference, with SE values taken from the \citet{Rakshit2020} catalog that appear to be consistently smaller than the X-ray-based ones, despite being in a regime of accretion rate and luminosity where the two methods are expected to be fully consistent. 
      
      A comparison with other catalogs of SDSS AGN indicates that this discrepancy is related to the specific way that the spectral decomposition and width measurements are performed by \citet{Rakshit2020}, which in turn causes the \mbh\ values to be systematically underestimated by a factor of 2.5. Once the SE \mbh\ values are recalibrated, then the only relevant discrepancies are observed for very highly accreting AGN, whose SE-based \mbh\ values are significantly overestimated, and for substantially absorbed AGN, whose SE-based \mbh\ values are instead underestimated.
      \item Using the X-ray-based \mbh\ values, we investigated various correlations and  confirmed strong positive correlations for the photon index $\Gamma$ vs. the Eddington ratio \laedd\ and for the X-ray bolometric correction vs. \laedd,  as well as for $\Gamma$ vs. the soft excess strength, in contrast with the results obtained using SE measurements by \citet{Laurenti2024}.
   \end{enumerate}

We end with a cautionary note and a speculation on the nature of LRDs. When using  SE-based \mbh\ provided in catalogs, one should keep in mind that not all catalogs are equally viable (for example, we have shown that the one from \citet{Rakshit2020} has \mbh\ values systematically underestimated by a factor of 2.5). Additionally, the use of quantities flagged for bad quality and \mbh\ values based on the \ion{C}{iv} line should be avoided, since they yield poor estimates of the virial mass.

As for the LRDs, our study confirms that the most extreme super-Eddington sources are very X-ray weak and characterized by steep photon indices, which may explain why LRDs are so difficult to detect by current X-ray observatories. Additionally, our work suggests that the  SE \mbh\ of sources accreting well above the Eddington level appear to be overestimated by about one order of magnitude. We can therefore speculate that LRDs (which are thought to be AGN in an early highly accreting phase) are likely less overmassive than currently thought and this may relax some of the constraints on BH seed models.

\begin{acknowledgements}
      Based on observations obtained with XMM-Newton, an ESA science mission with instruments and contributions directly funded by ESA Member States and NASA.
      We thank the referee for the constructive comments and suggestions that have improved the clarity of this paper.
      MG thanks Misty Bentz for useful discussions and Terrvon L. J. Kelley and Michael Young for their help with the soft excess analysis. 
\end{acknowledgements}

%
%

 \bibliographystyle{aa} 
 \bibliography{MonsterMBHs} 

\begin{appendix}
\section{X-ray scaling method in a nutshell} 
Here we briefly summarize the main characteristics of the X-ray scaling method, since its details have already been described at length in several papers \citep{Shaposhnikov2009,Gliozzi2011,Gliozzi2021,Williams2023}. 

This method lies on two pillars: 1) the luminosity of any BH accreting system is proportional to the \mbh, the accretion rate (in Eddington units) $\dot{m}$, and the radiative efficiency $\eta$ and 2) the photon index $\Gamma$ is a reliable indicator of the accretion state of any BH system. In other words, comparing BH systems with the same $\Gamma$ ensures that the systems are in the same spectral state and hence have roughly the same values of $\dot{m}$ and $\eta$.

In practice, this method determines the AGN \mbh\ by scaling up the dynamically constrained mass of a stellar mass BH system (the reference source) using two parameters of the \textsc{bmc} model: the spectral index $\alpha$ (where $\Gamma=\alpha+1$) and the normalization $N_{\mathrm{BMC}}$. The former ensures that the AGN and reference source are in the same spectral state and the latter computes the actual scaling process:

\begin{equation}\label{eq_Xscal1}
\left(\frac{N_{\mathrm{BMC,AGN}}}{N_{\mathrm{BMC,ref}}}\right) = \left(\frac{(\eta~ \dot{m}~  M_{\mathrm{BH}})_{\mathrm{AGN}}}{(\eta ~ \dot{m}~  M_{\mathrm{BH}})_{\mathrm{ref}}}\right) \cdot \left(\frac{d_{\mathrm{ref}}^2}{d_{\mathrm{AGN}}^2}\right)
= \left( \frac{ M_{\mathrm{BH,AGN}}} {M_{\mathrm{BH,ref}}}\right) \cdot \left(\frac{d_{\mathrm{ref}}^2}{d_{\mathrm{AGN}}^2}\right)
\end{equation}

\noindent{Solving for the AGN \mbh:}
\begin{equation}\label{eq_Xscal2}
M_{\mathrm{BH,AGN}}=M_{\mathrm{BH,ref}}  \cdot \left(\frac{N_{\mathrm{BMC,AGN}}}{N_{\mathrm{BMC,ref}}}\right)  \cdot \left(\frac{d_{\mathrm{AGN}}^2}{d_{\mathrm{ref}}^2}\right)
\end{equation}

\begin{figure} 
  \includegraphics[width=\columnwidth]{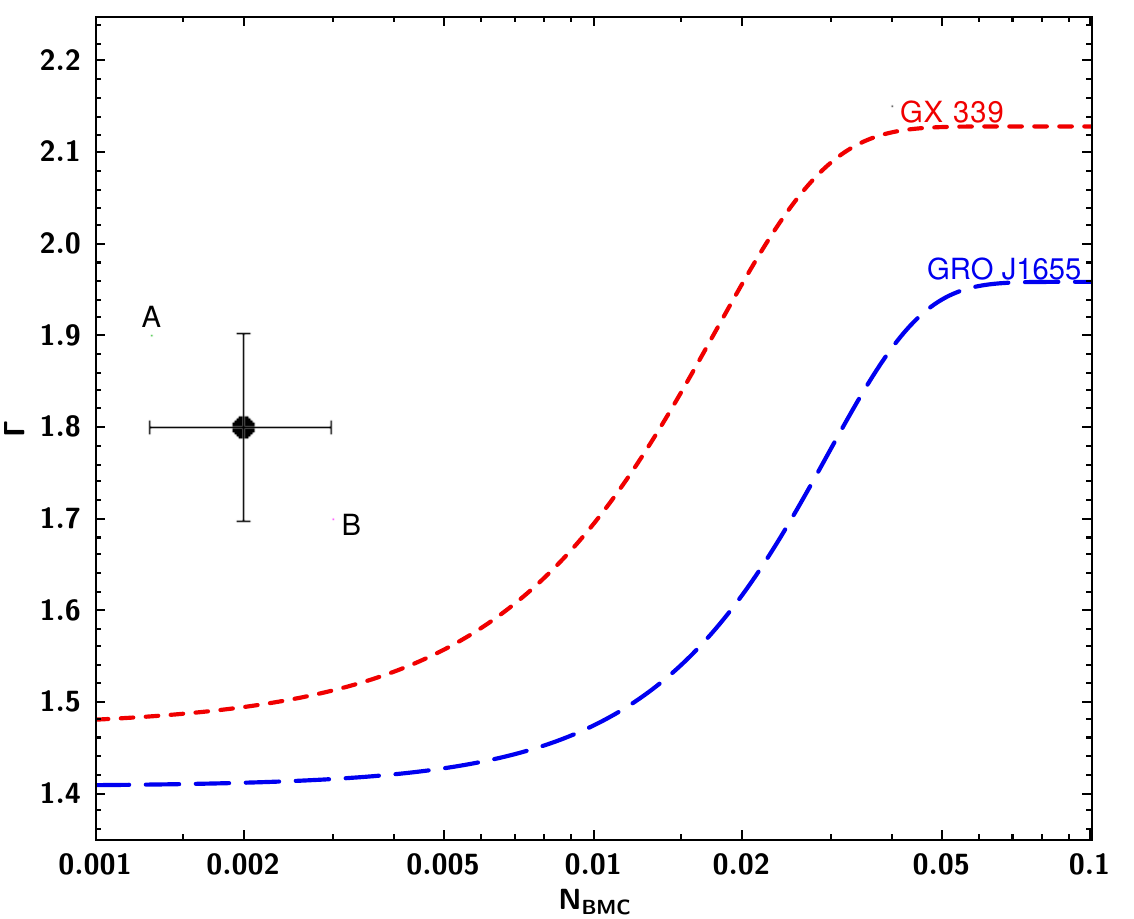} 
  \caption{$\Gamma$ - $N_\mathrm{BMC}$ diagram where the black point represents the AGN with its uncertainties (which have been increased for illustration purposes). The red short-dashed line indicates the spectral trend of the reference source GX 339-4. The blue long-dashed line describes the trend of the reference source GRO J1655-40.}
  \label{fig:GaNbmc}
\end{figure}

This process is illustrated by the $\Gamma$ - $N_\mathrm{BMC}$ diagram in Fig.~\ref{fig:GaNbmc}, where a generic AGN is compared to the two most reliable reference sources GX 339-4 (whose spectral transition is indicated by the red short-dashed line) and GRO J1655-40 (blue long-dashed line). 

The position of the AGN along the y-axis selects the corresponding spectral state in the reference source, whereas the separation along the x-axis yields the scaling of the \mbh. The larger the separation, the smaller the AGN mass; therefore, GRO J1655-40 consistently yields slightly lower values than GX 339-4. For $\Gamma$ values in the range viable for both reference sources, we compute the \mbh\ values for each reference source and take the average. 

The uncertainties on \mbh\ are computed accounting for the uncertainties on the spectral parameters: the distance between point $A$ (defined by the coordinates $N_\mathrm{BMC}-\mathrm{error}$ $\Gamma+\mathrm{error}$) and the GRO J1655-40 trend defines the minimum \mbh, whereas the distance between point $B$ ($N_\mathrm{BMC}+\mathrm{error}$ $\Gamma-\mathrm{error}$) and the GX 339-4 trend yields the maximum \mbh. We note that for sources with relatively flat photon indices ($\Gamma < 1.5$), only GRO J1655-40 is used, whereas for steeper sources ($\Gamma > 1.95$) we use only GX 339-4. Finally, for very steep sources ($\Gamma > 2.2$), we utilize a third reference source, XTE J1550-564, which was not included in Fig.~\ref{fig:GaNbmc} for the sake of clarity.

\section{Samples used in this paper}


\begin{table*}
  \caption{WISSH subsample}
  \label{tab:wissh}
  \centering
  \begin{tabular}{cccccccc}
    \hline
    \hline
    \noalign{\smallskip}
    Source & $z$ & N$_\mathrm{H}$ & $\Gamma$ & $N_\mathrm{BMC}$ & \mbh & $L_\mathrm{X}$ & $\lambda_\mathrm{Edd}$ \\
    & & ($10^{-22}$ cm$^{-2}$) & & & (\msun) & (erg s$^{-1}$) \\
    (1) & (2) & (3) & (4) & (5) & (6) & (7) & (8) \\
    \hline
    \noalign{\smallskip}
    SDSS 074711.14+273903.3 & 4.109 & $1.19_{-0.87}^{+1.05}$ & $1.62_{-0.13}^{+0.13}$ & $6.48_{-0.66}^{+0.79} \times 10^{-8}$ & $2.86_{-0.38}^{+0.53} \times 10^9$ & $1.03 \times 10^{45}$ & 0.55 \\
    \noalign{\smallskip}
    SDSS 090033.50+421547.0 & 3.294 & $1.43_{-0.18}^{+0.18}$ & $1.74_{-0.04}^{+0.04}$ & $2.49_{-0.05}^{+0.05} \times 10^{-6}$ & $5.83_{-1.60}^{+2.94} \times 10^{10}$ & $1.53 \times 10^{46}$ & 0.08 \\
    \noalign{\smallskip}
    SDSS 090423.37+130920.7 & 2.977 & $0.84_{-0.17}^{+0.18}$ & $2.08_{-0.04}^{+0.04}$ & $1.17_{-0.07}^{+0.08} \times 10^{-6}$ & $1.03_{-0.18}^{+0.17} \times 10^{10}$ & $7.36 \times 10^{45}$ &   0.45 \\
    \noalign{\smallskip}
    SDSS 094734.19+142116.9 & 3.031 &
    $2.12_{-0.30}^{+0.32}$ & $2.04_{-0.07}^{+0.07}$ & $5.52_{-0.34}^{+0.36} \times 10^{-7}$ & $5.85_{-0.74}^{+0.81} \times 10^9$ & $3.44 \times 10^{45}$ & 0.62 \\
    \noalign{\smallskip}
    SDSS 101447.18+430030.1 & 2.959 &
    $1.01_{-0.35}^{+0.39}$ & $1.84_{-0.12}^{+0.12}$ & $5.23_{-0.28}^{+0.29} \times 10^{-7}$ & $7.30_{-1.89}^{+2.80} \times 10^9$ & $2.29 \times 10^{45}$ & 0.75 \\
    \noalign{\smallskip}
    SDSS 102714.77+354317.4 & 3.118 &
    $< 0.21$ & $1.90_{-0.06}^{+0.06}$ & $7.14_{-0.34}^{+0.36} \times 10^{-7}$ & $9.70_{-2.79}^{+3.59} \times 10^9$ & $5.64 \times 10^{45}$ & 0.79 \\
    \noalign{\smallskip}
    SDSS 111038.63+483115.6 & 2.959 &
    $1.55_{-0.38}^{+0.41}$ & $1.99_{-0.08}^{+0.08}$ & $3.56_{-0.73}^{+0.93} \times 10^{-7}$ & $4.11_{-0.46}^{+0.55} \times 10^9$ & $2.18 \times 10^{45}$ & 1.21 \\
    \noalign{\smallskip}
    SDSS 125005.72+263107.5 & 2.044 &
    $< 0.09$ & $2.03_{-0.03}^{+0.03}$ & $3.56_{-0.08}^{+0.08} \times 10^{-6}$ & $1.49_{-0.18}^{+0.20} \times 10^{10}$ & $8.40 \times 10^{45}$ & 0.42 \\
    \noalign{\smallskip}
    SDSS 142656.18+602550.8 & 3.197 &
    $2.65_{-0.36}^{+0.39}$ & $1.95_{-0.08}^{+0.08}$ & $9.78_{-0.44}^{+0.46} \times 10^{-7}$ & $1.18_{-0.16}^{+0.52} \times 10^{10}$ & $4.70 \times 10^{45}$ & 3.28 \\
    \noalign{\smallskip}
    SDSS 154938.72+124509.1 & 2.386 &
    $4.28_{-0.37}^{+0.40}$ & $2.03_{-0.07}^{+0.07}$ & $5.16_{-1.0}^{+1.3} \times 10^{-7}$ & $3.25_{-0.38}^{+0.44} \times 10^9$ & $1.43 \times 10^{45}$ & 1.53 \\
    \noalign{\smallskip}
    SDSS 170100.60+641209.3 & 2.741 &
    $0.95_{-0.14}^{+0.15}$ & $2.21_{-0.04}^{+0.04}$ & $1.45_{-0.04}^{+0.04} \times 10^{-6}$ & $8.01_{-0.50}^{+0.80} \times 10^9$ & $7.63 \times 10^{45}$ & 0.98 \\
    \noalign{\smallskip}
    SDSS 212329.46-005052.9 & 2.269 &
    $2.46_{-0.26}^{+0.27}$ & $1.99_{-0.07}^{+0.07}$ & $6.43_{-1.1}^{+1.3} \times 10^{-7}$ & $3.87_{-0.43}^{+0.52} \times 10^9$ & $2.36 \times 10^{45}$ & 1.04 \\
    \noalign{\smallskip}
    \hline
  \end{tabular}
  \tablefoot{
  Columns: 1 = source name, 2 = redshift, 3 = intrinsic column density , 4 = photon index, 5 = BMC normalization, 6 = black hole mass determined with the X-ray scaling method, 7 = 2--10 keV luminosity, 8 = Eddington ratio $\lambda_\mathrm{Edd}=L_\mathrm{bol}/L_\mathrm{Edd}$.
  }
\end{table*}

\begin{table*}
 \caption{Auxiliary table for Table~\ref{tab:xhess}}
  \label{tab:aux}
  \centering
  \begin{tabular}{rlrlrl}
    \toprule
    \midrule
    Number & Name & Number & Name & Number & Name \\
    \midrule
    1 & SDSS J094610.71+095226.3 & 23 & SDSS J002209.69+013213.0 
      & 43 & SDSS J110312.93+414154.9 \\
    2 & SDSS J095847.88+690532.7 & 24 & SDSS J014634.38-093014.3 
      & 44 & SDSS J112306.33+013749.6 \\
    3 & SDSS J122549.87+332454.9 & 25 & SDSS J014904.48+125746.2
      & 45 & SDSS J112317.51+051804.0 \\
    4 & SDSS J113233.55+273956.3 & 26 & SDSS J015828.31-014810.0 
      & 47 & SDSS J112818.49+240217.4 \\
    5 & SDSS J130048.10+282320.6 & 28 & SDSS J022039.48-030820.3 
      & 48 & SDSS J120734.62+150643.7 \\
    6 & SDSS J172255.24+320307.5 & 29 & SDSS J024651.91-005930.9 
      & 49 & SDSS J120858.01+454035.4 \\
    9 & SDSS J021702.01+015352.0 & 31 & SDSS J081014.48+280337.1 
      & 50 & SDSS J124615.77+673032.7 \\
    11 & SDSS J154530.23+484608.9 & 32 & SDSS J081331.28+254503.0
       & 51 & SDSS J125005.72+263107.5 \\
    13 & SDSS J100402.61+285535.3 & 33 & SDSS J083850.15+261105.4 
       & 52 & SDSS J125216.58+052737.7 \\
    14 & SDSS J221715.18+002615.0 & 34 & SDSS J084153.99+194303.1 
       & 53 & SDSS J130112.91+590206.6 \\
    15 & SDSS J221738.41+001206.5 & 36 & SDSS J090033.50+421547.0 
       & 54 & SDSS J132654.95-000530.1 \\
    17 & SDSS J074545.01+392700.9 & 37 & SDSS J092247.03+512038.0 
       & 56 & SDSS J135306.34+113804.7 \\
    18 & SDSS J114229.22+264012.4 & 38 & SDSS J092943.41+004127.3 
       & 57 & SDSS J144741.76-020339.1 \\
    19 & SDSS J123034.20+073305.3 & 39 & SDSS J093922.89+370944.0 
       & 58 & SDSS J161434.67+470420.0 \\
    20 & SDSS J140621.89+222346.5 & 40 & SDSS J094033.75+462315.0 
       & 60 & SDSS J163201.11+373749.9 \\
    21 & SDSS J145108.76+270926.9 & 41 & SDSS J103928.14+392342.1
       & 61 & SDSS J223607.68+134355.3 \\
    22 & SDSS J233317.38-002303.4 & 42 & SDSS J110035.00+101027.4
       & & \\
    \bottomrule
  \end{tabular}
\end{table*}
    
\longtab[3]{
\begin{longtable}{ccccccccc}
  \caption{X-HESS subsample}
  \label{tab:xhess} \\
  \toprule
  \midrule
  Source no. & $z$ & N$_\mathrm{H}$ & $\Gamma$ & $N_\mathrm{BMC}$ & \mbh & $L_\mathrm{X}$ & $\lambda_\mathrm{Edd}$ & $\mathit{SX3}$ \\
  & & ($10^{-22}$ cm$^{-2}$) & & & (\msun) & (erg s$^{-1}$) & & \\
  (1) & (2) & (3) & (4) & (5) & (6) & (7) & (8) & (9) \\
  \midrule
  \endfirsthead
  \caption{continued.} \\
  \toprule
  \midrule
  Source no. & $z$ & $N_H$ & $\Gamma$ & $N_\mathrm{BMC}$ & \mbh & $L_\mathrm{X}$ & $\lambda_\mathrm{Edd}$ & $\mathit{SX3}$ \\
  & & ($10^{-22}$ cm$^{-2}$) & & & (\msun) & (erg s$^{-1}$) & & \\
  (1) & (2) & (3) & (4) & (5) & (6) & (7) & (8) & (9) \\
  \midrule
  \endhead
  \bottomrule
  \endfoot
  \bottomrule
  \endlastfoot
  \noalign{\smallskip}
  1 & 0.6975 & $< 0.23$ & $2.02_{-0.13}^{+0.14}$ & $8.36_{-0.46}^{+0.49} \times 10^{-7}$ & $2.50_{-0.30}^{+0.34} \times 10^8$ & $7.76 \times 10^{43}$ & $0.12$ & $3.41 \pm {0.50} \times 10^{-3}$ \\
  \noalign{\smallskip}
  2 & 1.2879 & $< 0.15$ & $2.04_{-0.04}^{+0.04}$ & $1.13_{-0.02}^{+0.02} \times 10^{-6}$ & $1.46_{-0.18}^{+0.20} \times 10^9$ & $9.99 \times 10^{44}$ & $0.67$ & \ldots \\
  \noalign{\smallskip}
  3 & 1.1421 & $1.85_{-0.79}^{+0.99}$ & $1.68_{-0.28}^{+0.30}$ & $6.35_{-2.0}^{+4.2} \times 10^{-8}$ & $1.31_{-0.40}^{+0.89} \times 10^8$ & $5.79 \times 10^{43}$ & $0.29$ & $1.13 \pm {0.66} \times 10^{-1}$ \\
  \noalign{\smallskip}
  4 & 0.6812 & $< 0.29$ & $2.43_{-0.22}^{+0.24}$ & $3.86_{-0.52}^{+0.59} \times 10^{-6}$ & $4.54_{-1.10}^{+1.73} \times 10^8$ & $5.16 \times 10^{43}$ & $0.13$ & \dots \\
  \noalign{\smallskip}
  5 & 1.9233 & $4.62_{-0.71}^{+0.83}$ & $2.67_{-0.27}^{+0.31}$ & $1.80_{-0.25}^{+0.27} \times 10^{-7}$ & $1.58_{-0.39}^{+0.60} \times 10^8$ & $6.58 \times 10^{44}$ & $9.70$ & \dots \\
  \noalign{\smallskip}
  6 & 0.2752 & $< 0.20$ & $1.53_{-0.09}^{+0.08}$ & $6.42_{-0.20}^{+0.20} \times 10^{-6}$ & $5.52_{-1.04}^{+1.91} \times 10^8$ & $7.19 \times 10^{43}$ & $0.01$ & \dots \\
  \noalign{\smallskip}
  9 & 0.7227 & $0.36_{-0.30}^{+0.33}$ & $1.71_{-0.12}^{+0.13}$ & $1.61_{-0.20}^{+0.26} \times 10^{-7}$ & $9.68_{-2.77}^{+5.56} \times 10^7$ & $6.11 \times 10^{43}$ & $0.08$ & \ldots \\
  \noalign{\smallskip}
  11 & 0.3998 & $< 0.15$ & $2.45_{-0.13}^{+0.13}$ & $5.05_{-0.32}^{+0.34} \times 10^{-6}$ & $1.57_{-0.38}^{+0.60} \times 10^8$ & $1.84 \times 10^{44}$ & $1.23$ & $1.06 \pm {0.33} \times 10^{-2}$ \\
  \noalign{\smallskip}
  13 & 0.3291 & $< 0.07$ & $2.07_{-0.04}^{+0.04}$ & $7.95_{-0.14}^{+0.15} \times 10^{-6}$ & $3.33_{-0.52}^{+0.51} \times 10^8$ & $1.96 \times 10^{44}$ & $0.58$ & $6.35 \pm {0.98} \times 10^{-3}$ \\
  \noalign{\smallskip}
  14 & 0.7530 & $3.26_{-0.20}^{+0.21}$ & $2.05_{-0.14}^{+0.15}$ & $1.76_{-0.07}^{+0.08} \times 10^{-6}$ & $5.77_{-0.77}^{+0.82} \times 10^8$ & $1.82 \times 10^{44}$ & 0.04 & \ldots \\
  \noalign{\smallskip}
  15 & 1.1232 & $< 0.34$ & $1.70_{-0.17}^{+0.18}$ & $1.19_{-0.08}^{+0.08} \times 10^{-7}$ & $2.20_{-0.64}^{+1.33} \times 10^8$ & $8.20 \times 10^{43}$ & $0.18$ & \ldots \\
  \noalign{\smallskip}
  17 & 1.6279 & $0.54_{-0.34}^{+0.37}$ & $1.97_{-0.12}^{+0.13}$ & $4.34_{-0.37}^{+0.42} \times 10^{-7}$ & $1.22_{-0.13}^{+0.17} \times 10^9$ & $6.07 \times 10^{44}$ & $0.20$ & \ldots \\
  \noalign{\smallskip}
  18 & 1.6771 & $< 0.47$ & $2.03_{-0.09}^{+0.09}$ & $8.33_{-0.51}^{+0.06} \times 10^{-7}$ & $2.12_{-0.26}^{+0.29} \times 10^9$ & $1.30 \times 10^{45}$ & $0.11$ & \ldots \\
  \noalign{\smallskip}
  19 & 1.8158 & $< 0.27$ & $1.87_{-0.10}^{+0.10}$ & $1.08_{-0.07}^{+0.08} \times 10^{-6}$ & $4.27_{-1.14}^{+1.58} \times 10^9$ & $2.30 \times 10^{45}$ & $0.29$ & \ldots \\
  \noalign{\smallskip}
  20 & 0.0983 & $3.75_{-0.16}^{+0.17}$ & $1.56_{-0.12}^{+0.10}$ & $2.38_{-0.09}^{+0.10} \times 10^{-5}$ & $1.87_{-0.31}^{+0.49} \times 10^8$ & $4.63 \times 10^{42}$ & $0.03$ & \ldots \\
  \noalign{\smallskip}
  21 & 0.0645 & $4.43_{-0.03}^{+0.03}$ & $2.50_{-0.01}^{+0.01}$ & $4.02_{-0.05}^{+0.05} \times 10^{-4}$ & $2.02_{-0.49}^{+0.76} \times 10^8$ & $1.25 \times 10^{43}$ & $0.08$ & \ldots \\
  \noalign{\smallskip}
  22 & 0.5130 & $< 0.35$ & $2.03_{-0.17}^{+0.18}$ & $4.07_{-1.8}^{+3.4} \times 10^{-6}$ & $5.55_{-0.68}^{+0.76} \times 10^8$ & $2.68 \times 10^{44}$ & $0.01$ & \ldots \\
  \noalign{\smallskip}
  23 & 1.8826 & $< 0.80$ & $2.01_{-0.16}^{+0.16}$ & $2.41_{-0.86}^{+1.4} \times 10^{-7}$ & $8.69_{-1.00}^{+1.17} \times 10^8$ & $5.24 \times 10^{44}$ & $0.44$ & \ldots \\
  \noalign{\smallskip}
  24 & 0.3934 & $< 0.28$ & $1.75_{-0.18}^{+0.18}$ & $1.56_{-0.08}^{+0.08} \times 10^{-6}$ & $1.92_{-0.52}^{+0.93} \times 10^8$ & $5.17 \times 10^{43}$ & $0.06$ & $2.39 \pm {0.91} \times 10^{-3}$ \\
  \noalign{\smallskip}
  25 & 0.7305 & $2.43_{-0.39}^{+0.43}$ & $2.08_{-0.16}^{+0.17}$ & $5.10_{-0.70}^{+0.84} \times 10^{-7}$ & $1.39_{-0.25}^{+0.22} \times 10^8$ & $1.01 \times 10^{44}$ & $0.18$ & $2.95 \pm {0.63} \times 10^{-1}$ \\
  \noalign{\smallskip}
  26 & 1.7719 & $0.88_{-0.16}^{+0.17}$ & $2.34_{-0.07}^{+0.07}$ & $1.51_{-0.11}^{+0.13} \times 10^{-6}$ & $2.25_{-0.55}^{+0.87} \times 10^9$ & $2.17 \times 10^{45}$ & $0.54$ & \ldots \\
  \noalign{\smallskip}
  28 & 0.4519 & $1.49_{-0.31}^{+0.34}$ & $2.93_{-0.29}^{+0.33}$ & $1.85_{-0.16}^{+0.17} \times 10^{-6}$ & $1.17_{-0.56}^{+0.72} \times 10^7$ & $8.31 \times 10^{43}$ & $3.30$ & $4.61 \pm {2.55} \times 10^{-1}$ \\
  \noalign{\smallskip}
  29 & 0.4675 & $< 0.23$ & $1.67_{-0.15}^{+0.15}$ & $4.73_{-0.22}^{+0.22} \times 10^{-6}$ & $1.12_{-0.35}^{+0.81} \times 10^9$ & $2.18 \times 10^{44}$ & $0.17$ & $1.03 \pm {0.54} \times 10^{-3}$ \\
  \noalign{\smallskip}
  31 & 0.8209 & $< 0.13$ & $1.66_{-0.08}^{+0.08}$ & $3.69_{-0.30}^{+0.37} \times 10^{-6}$ & $3.57_{-1.13}^{+2.78} \times 10^9$ & $1.10 \times 10^{45}$ & $0.07$ & \ldots \\
  \noalign{\smallskip}
  32 & 1.5099 & $0.43_{-0.06}^{+0.06}$ & $2.00_{-0.02}^{+0.02}$ & $3.92_{-0.05}^{+0.05} \times 10^{-6}$ & $7.12_{-1.86}^{+3.02} \times 10^9$ & $4.35 \times 10^{45}$ & $0.86$ & $5.26 \pm {1.85} \times 10^{-3}$ \\
  \noalign{\smallskip}
  33 & 1.6139 & $< 0.20$ & $1.94_{-0.07}^{+0.07}$ & $1.20_{-0.04}^{+0.04} \times 10^{-6}$ & $2.83_{-0.39}^{+1.15} \times 10^9$ & $1.69 \times 10^{45}$ & $1.08$ & \ldots \\
  \noalign{\smallskip}
  34 & 0.4786 & $< 0.30$ & $1.57_{-0.18}^{+0.19}$ & $5.59_{-0.39}^{+0.48} \times 10^{-7}$ & $1.50_{-0.24}^{+0.37} \times 10^8$ & $4.80 \times 10^{43}$ & $0.06$ & \ldots \\
  \noalign{\smallskip}
  36 & 3.2954 & $1.44_{-0.18}^{+0.18}$ & $1.74_{-0.04}^{+0.04}$ & $2.50_{-0.05}^{+0.05} \times 10^{-6}$ & $5.86_{-1.61}^{+2.95} \times 10^{10}$ & $1.53 \times 10^{46}$ & $0.17$ & \ldots \\
  \noalign{\smallskip}
  37 & 0.1597 & $< 0.10$ & $2.14_{-0.06}^{+0.06}$ & $1.04_{-0.04}^{+0.04} \times 10^{-5}$ & $7.06_{-1.79}^{+2.88} \times 10^7$ & $2.74 \times 10^{43}$ & $0.17$ & $7.52 \pm {2.49} \times 10^{-3}$ \\
  \noalign{\smallskip}
  38 & 0.5868 & $0.52_{-0.22}^{+0.24}$ & $2.41_{-0.14}^{+0.15}$ & $1.74_{-0.19}^{+0.22} \times 10^{-6}$ & $1.47_{-0.36}^{+0.56} \times 10^8$ & $1.55 \times 10^{44}$ & $1.65$ & $9.53 \pm {3.10} \times 10^{-3}$ \\
  \noalign{\smallskip}
  39 & 0.1861 & $< 0.32$ & $2.46_{-0.19}^{+0.20}$ & $4.14_{-0.36}^{+0.40} \times 10^{-6}$ & $2.18_{-0.53}^{+0.83} \times 10^7$ & $2.43 \times 10^{43}$ & $0.56$ & $2.54 \pm {0.80} \times 10^{-2}$ \\
  \noalign{\smallskip}
  40 & 0.6966 & $< 0.25$ & $2.30_{-0.15}^{+0.16}$ & $8.26_{-0.91}^{+1.0} \times 10^{-7}$ & $1.31_{-0.32}^{+0.51} \times 10^8$ & $1.47 \times 10^{44}$ & $1.47$ & $2.72 \pm {0.88} \times 10^{-2}$ \\
  \noalign{\smallskip}
  41 & 1.0115 & $1.86_{-0.10}^{+0.10}$ & $1.88_{-0.04}^{+0.04}$ & $1.30_{-0.03}^{+0.03} \times 10^{-5}$ & $1.17_{-0.32}^{+0.43} \times 10^{10}$ & $2.34 \times 10^{45}$ & $0.06$ & \ldots \\
  \noalign{\smallskip}
  42 & 1.5877 & $< 0.47$ & $2.07_{-0.12}^{+0.12}$ & $6.69_{-0.67}^{+0.76} \times 10^{-7}$ & $1.31_{-0.21}^{+0.20} \times 10^9$ & $9.33 \times 10^{44}$ & $0.37$ & \ldots \\
  \noalign{\smallskip}
  43 & 0.4016 & $3.00_{-0.43}^{+0.59}$ & $1.53_{-0.27}^{+0.20}$ & $7.04_{-0.63}^{+0.73} \times 10^{-6}$ & $1.47_{-0.28}^{+0.51} \times 10^9$ & $4.64 \times 10^{43}$ & $0.08$ & \ldots \\
  \noalign{\smallskip}
  44 & 0.6957 & $1.23_{-0.31}^{+0.34}$ & $2.26_{-0.21}^{+0.23}$ & $4.97_{-0.86}^{+1.1} \times 10^{-7}$ & $8.49_{-2.11}^{+3.35} \times 10^7$ & $8.76 \times 10^{43}$ & $0.45$ & \ldots \\
  \noalign{\smallskip}
  45 & 0.9986 & $2.97_{-0.27}^{+0.30}$ & $1.59_{-0.21}^{+0.16}$ & $6.48_{-0.32}^{+0.39} \times 10^{-6}$ & $9.82_{-1.43}^{+2.12} \times 10^{9}$ & $1.04 \times 10^{45}$ & $0.06$ & \ldots \\
  \noalign{\smallskip}
  47 & 1.6071 & $< 0.16$ & $1.76_{-0.05}^{+0.05}$ & $7.99_{-0.27}^{+0.28} \times 10^{-7}$ & $3.07_{-0.82}^{+1.44} \times 10^9$ & $1.36 \times 10^{45}$ & $0.20$ & \ldots \\
  \noalign{\smallskip}
  48 & 0.7507 & $< 0.17$ & $1.88_{-0.12}^{+0.12}$ & $1.18_{-0.06}^{+0.07} \times 10^{-6}$ & $5.05_{-1.37}^{+1.86} \times 10^8$ & $2.90 \times 10^{44}$ & $0.48$ & $2.19 \pm {0.76} \times 10^{-3}$ \\
  \noalign{\smallskip}
  49 & 1.1624 & $1.24_{-0.21}^{+0.22}$ & $1.78_{-0.13}^{+0.13}$ & $2.33_{-0.09}^{+0.10} \times 10^{-6}$ & $3.79_{-1.00}^{+1.66} \times 10^9$ & $1.50 \times 10^{45}$ & $0.81$ & $5.15 \pm {1.65} \times 10^{-2}$ \\
  \noalign{\smallskip}
  50 & 1.7656 & $< 0.24$ & $1.96_{-0.10}^{+0.11}$ & $6.26_{-0.41}^{+0.45} \times 10^{-7}$ & $2.20_{-0.24}^{+0.30} \times 10^9$ & $1.20 \times 10^{45}$ & $0.28$ & \ldots \\
  \noalign{\smallskip}
  51 & 2.0441 & $< 0.09$ & $2.03_{-0.03}^{+0.03}$ & $3.38_{-0.07}^{+0.08} \times 10^{-6}$ & $1.42_{-0.17}^{+0.19} \times 10^{10}$ & $8.71 \times 10^{45}$ & $0.86$ & \ldots \\
  \noalign{\smallskip}
  52 & 1.9010 & $0.96_{-0.19}^{+0.19}$ & $2.29_{-0.07}^{+0.07}$ & $1.14_{-0.07}^{+0.07} \times 10^{-6}$ & $2.24_{-0.55}^{+0.88} \times 10^9$ & $2.27 \times 10^{45}$ & $1.37$ & $7.41 \pm {2.67} \times 10^{-2}$ \\
  \noalign{\smallskip}
  53 & 0.4762 & $0.10_{-0.09}^{+0.09}$ & $2.31_{-0.07}^{+0.07}$ & $7.08_{-0.36}^{+0.39} \times 10^{-6}$ & $4.36_{-1.08}^{+1.70} \times 10^8$ & $3.61 \times 10^{44}$ & $0.88$ & $1.27 \pm {0.41} \times 10^{-2}$ \\
  \noalign{\smallskip}
  54 & 3.3070 & $5.06_{-0.67}^{+0.75}$ & $1.76_{-0.08}^{+0.09}$ & $2.77_{-0.20}^{+0.23} \times 10^{-7}$ & $6.19_{-1.66}^{+2.90} \times 10^9$ & $1.98 \times 10^{45}$ & $2.48$ & \ldots \\
  \noalign{\smallskip}
  56 & 1.6282 & $1.66_{-0.11}^{+0.11}$ & $2.38_{-0.05}^{+0.05}$ & $1.88_{-0.08}^{+0.08} \times 10^{-6}$ & $2.12_{-0.52}^{+0.82} \times 10^9$ & $2.19 \times 10^{45}$ & $0.91$ & $1.40 \pm {0.45} \times 10^{-1}$ \\
  \noalign{\smallskip}
  57 & 1.4273 & $0.43_{-0.24}^{+0.25}$ & $2.70_{-0.14}^{+0.15}$ & $4.68_{-0.33}^{+0.35} \times 10^{-7}$ & $1.79_{-0.44}^{+0.68} \times 10^8$ & $4.18 \times 10^{44}$ & $1.71$ & $4.02 \pm {1.40} \times 10^{-2}$ \\
  \noalign{\smallskip}
  58 & 1.8645 & $< 0.08$ & $1.76_{-0.01}^{+0.01}$ & $3.74_{-0.03}^{+0.03} \times 10^{-6}$ & $2.06_{-0.55}^{+0.96} \times 10^{10}$ & $9.00 \times 10^{45}$ & $0.24$ & $8.45 \pm {3.18} \times 10^{-4}$ \\
  \noalign{\smallskip}
  60 & 1.4757 & $2.49_{-0.16}^{+0.17}$ & $1.97_{-0.07}^{+0.07}$ & $5.23_{-0.21}^{+0.21} \times 10^{-6}$ & $1.15_{-0.13}^{+0.16} \times 10^{10}$ & $1.95 \times 10^{45}$ & $0.27$ & \ldots \\
  \noalign{\smallskip}
  61 & 0.3257 & $< 0.15$ & $2.13_{-0.14}^{+0.15}$ & $7.32_{-0.22}^{+0.22} \times 10^{-6}$ & $2.50_{-0.64}^{+1.02} \times 10^8$ & $1.66 \times 10^{44}$ & $0.49$ & $6.03 \pm {2.04} \times 10^{-3}$ \\
\end{longtable}
\flushleft
Columns: 1 = source number, 2 = redshift, 3 = intrinsic column density , 4 = photon index, 5 = BMC normalization, 6 = black hole mass determined with the X-ray scaling method, 7 = 2--10 keV luminosity, 8 = Eddington ratio $\lambda_\mathrm{Edd}=L_\mathrm{bol}/L_\mathrm{Edd}$, 9 = soft excess strength = $L_{\rm bb, 0.5-2\,keV}/L_{\rm Edd}$.
} 

\end{appendix}

\end{document}